\documentclass[review,number,sort&compress]{elsarticle}

\usepackage{color}
\usepackage{amssymb}
\usepackage{graphicx}
\journal{Nuclear Instruments and Methods A}

\begin{document}

\begin{frontmatter}
\title{The Central Drift Chamber for GlueX}
\author[CMU]{N.~S.~Jarvis}
\author[CMU]{C.~A.~Meyer}
\author[Jlab]{B.~Zihlmann}
\author[CMU]{M.~Staib}
\author[CMU]{A.~Austregesilo}
\author[Jlab]{F.~Barbosa}
\author[Jlab]{C.~Dickover}
\author[Jlab]{V.~Razmyslovich}
\author[Jlab]{S.~Taylor}
\author[CSIRO]{Y.~Van~Haarlem}
\author[IU]{G.~Visser}
\author[Jlab]{T.~Whitlatch}

\address[CMU]{Department of Physics, Carnegie Mellon University, Pittsburgh, Pennsylvania~15213, USA}
\address[Jlab]{Thomas Jefferson National Accelerator Facility, Newport News, Virginia~23606, USA}
\address[CSIRO]{Commonwealth Scientific and Industrial Research Organisation, Lucas Heights, NSW~2234, Australia}
\address[IU]{Department of Physics, Indiana University, Bloomington, Indiana~47405, USA}

\begin{abstract}
The Central Drift Chamber is a straw-tube wire chamber of cylindrical structure located surrounding the target inside the bore of the GlueX spectrometer solenoid. Its purpose is to detect and track charged particles with momenta as low as 0.25\, GeV/c and to identify low-momentum protons via energy loss. The construction of the detector is described and its operation and calibration are discussed in detail. The design goal of 150~$\mu$m in position resolution has been reached.
\end{abstract}

\begin{keyword}
29.40.Gx \sep 29.40.Cs 
\end{keyword}

\end{frontmatter}


\section{\label{Intro}Introduction}
The GlueX Central Drift Chamber (CDC) is part of the GlueX experiment in Hall D at Jefferson Lab in Newport News, VA. This experiment~\cite{gluex-ref,Dugger:2012qra,AlekSejevs:2013mkl,Dugger:2014xaa} aims to study the gluonic field in Quantum Chromodynamics (QCD) by searching for hybrid mesons with gluonic degrees of freedom and exotic quantum numbers~\cite{Meyer:2010ku,Meyer:2015eta}. Such states are expected in the mass range within reach for GlueX~\cite{Dudek:2013yja}. For the GlueX experiment, 12~GeV electrons from the Continuous Electron Beam Accelerator Facility are used to generate a high energy polarized photon beam through coherent bremsstrahlung from a thin diamond crystal. The highly collimated photon beam impinges on a liquid hydrogen target to produce various hadronic states that immediately decay into long-lived neutral and charged particles. Various detector systems including the CDC are used together to reconstruct the initial hadronic state based on the measurements of the decay particles.

The CDC is a cylindrical straw-tube drift chamber situated within the upstream end of the GlueX spectrometer solenoid. It surrounds an extended liquid hydrogen target and a `start counter' barrel scintillator and it is designed to track charged particles fully by providing position and energy loss measurements. The average position resolution of each straw must be about 150~$\mu$m to guarantee determination of the charged particles' momenta with a resolution of 2\% or better.

Prior  to  construction  of  the  CDC,  the  proposed  materials were  evaluated and used to build two  prototype  chambers with shorter straw-tubes and a sector with full length straw-tubes. These  prototypes were then  used  to  study  chamber characteristics and performance, including radiation resistance and choice of gas mix, reported in~\cite{VanHaarlem:2010yq}. This paper reports on the construction and performance of the CDC in GlueX, after several years of operational experience.

\section{\label{sec:cdc:construction}Construction}
\subsection{Design Criteria}
The materials and dimensions of the CDC were determined as part of a global design of the entire GlueX detector with the goal of providing hermetic coverage with similar resolution for both charged particles and photons. With regard to charged particles, an optimization was made between the relative lengths of the central tracking region around the target and the planar tracking volume downstream of the target in the solenoid. This study recommended lengths of approximately $1500$~mm and $2500$~mm for the central and planar tracking regions, respectively. The inner radius of the CDC was set by the size of the liquid hydrogen target and the start counter~\cite{Pooser:2019rhu}, while the outer radius was set by the bore of the solenoid and the depth needed for the barrel calorimeter~\cite{Beattie:2018xsk}. Materials and thicknesses were chosen to minimize scattering in the central and downstream regions of the chamber while maintaining structural integrity. This led to the choice of aluminum in the upstream end of the chamber with a carbon fiber endplate and plastic components in the downstream end of the chamber.

The size of the tubes was chosen to provide sufficient tracking layers in the central volume, while also maximizing the tracking regions more than $1.5$~mm away from both the central wire and the tube wall. The combination of straight and stereo straws was chosen to provide the  momentum resolution required, while also yielding the best resolution along the lengths of the straws. The $6^{\circ}$ stereo angles were chosen to be as large as possible while still limiting the radial dead space in the 4 transitions between straight and stereo blocks of straws. The desired position resolution of 150~$\mu$m together with the stereo angles of $\pm6^\circ$ at the transition points gave a polar angle resolution of approximately 0.5$^\circ$.

\subsection{\label{sec:cdc:overview} Overview}
\begin{figure}[ht!]\centering
\includegraphics[width=0.45\textwidth]{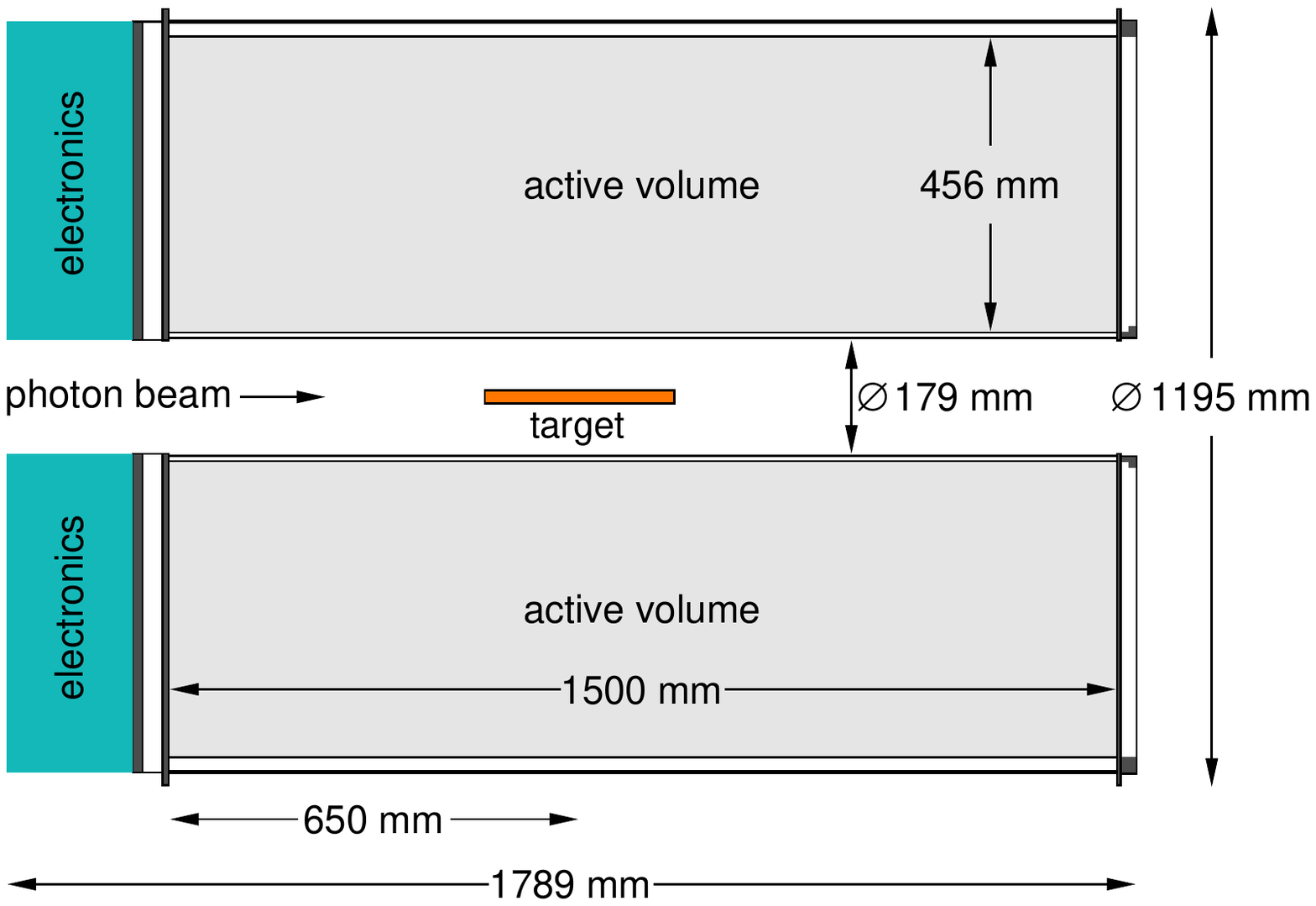}
\caption[]{\label{fig:xsection}Drawing of an axial cross section through the CDC. The liquid hydrogen volume in the target is 292~mm long and 16~mm in diameter.} 
\end{figure}

The CDC contains 3522 straw-tubes of diameter 16~mm arranged in 28~layers, located in a cylindrical volume with a length of 1500~mm, an inner radius of 99~mm and an outer radius of 555~mm. An axial cross-section through the chamber is shown in the diagram in Fig.~\ref{fig:xsection}. Each tube contains an anode wire of 20~$\mu$m diameter gold-plated tungsten. A layer of aluminum on the inside wall of the tube forms the cathode.  The tubes contribute structural rigidity to the assembly, support the tension of the wires, provide a uniform electric field and also prevent the wires from making contact with their neighbors, which would cause extensive electrical shorts in the event that one should break. 

\begin{figure}[ht!]\centering
\includegraphics[width=0.45\textwidth]{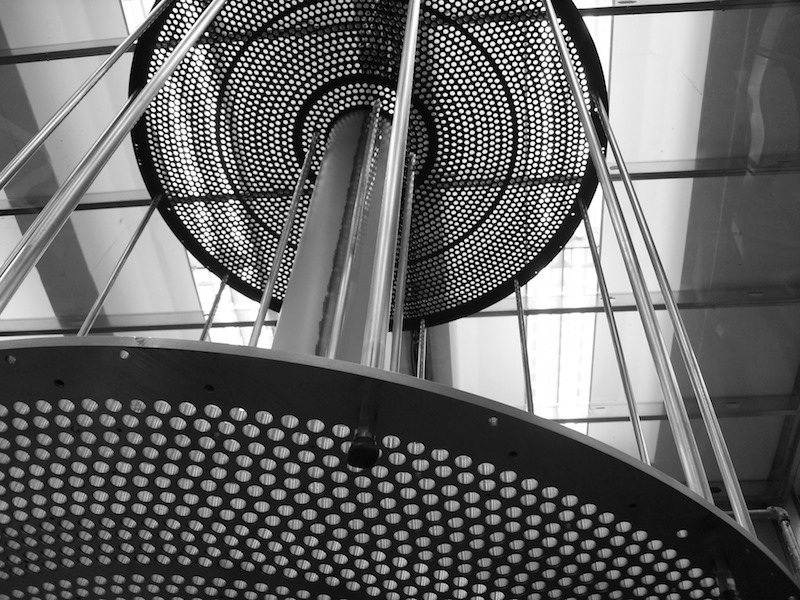}
\caption[]{\label{fig:cdc_photo_frame}CDC frame prior to the installation of the straw-tubes.} 
\end{figure}

The tracking volume is enclosed by an inner shell of G-10, an outer shell of aluminum, a carbon fiber endplate at the downstream (forward) end and an aluminum endplate at the upstream end. The endplates are linked by 12 aluminum support rods that were bolted into place to maintain the relative location of the endplates after alignment. Figure~\ref{fig:cdc_photo_frame} shows the endplates, inner shell, and support rods before the straws were installed.  The holes in the endplates were milled precisely to position the ends of the straws correctly.

There is a cylindrical gas plenum outside each endplate. The upstream plenum has polycarbonate sidewalls and a polycarbonate endplate, while the downstream plenum has ROHACELL\footnote{www.rohacell.com} sidewalls and a thin endwall of Mylar\footnote{www.mylar.com} film,  aluminized on both sides.  The inner and outer shells are sealed along their seams and where they meet the endplates, forming another plenum around the straws. The dimensions of the CDC are given in Table~\ref{tab:cdc_geometry}.  
\begin{table}[ht]\centering
\caption[]{\label{tab:cdc_geometry}The geometry of the CDC.}
\begin{tabular}{lr} \hline
Active volume inner radius    & 99.2~mm \\
Active volume outer radius    & 555.4~mm\\
Active length                 & 1500.0~mm \\
Chamber assembly inner radius & 87.5~mm \\
Chamber assembly outer radius & 597.4~mm \\
Upstream gas plenum length  & 31.8~mm \\
Downstream gas plenum length & 25.4~mm \\
Thickness of 28 straws, mylar & 2.22\,\% Rad.Length\\
Thickness of 28 straws, gas & 0.34\,\% Rad.Length\\
Thickness of downstream endplate & 2.14\,\% Rad.Length\\
\hline
\end{tabular}
\end{table}

The electronics are mounted on standoffs on the polycarbonate endplate. Signal cables pass through holes in that endplate (sealed with a threaded bushing and an O-ring) and then through the gas plenum to reach the crimp pins which hold the anode wires in place.

The CDC construction took place inside a class 2000 cleanroom at Carnegie Mellon University. Some initial testing took place there before it was transported to Jefferson Lab in October 2013.

\subsection{\label{sec:cdc:straws} Straw-tubes}
The straw-tubes were manufactured by Lamina Dielectrics\footnote{www.lamina.uk.com} from four layers of Mylar tape wound into a tube.  The innermost layer of tape has 100~nm of aluminum vapor-deposited onto the side that faces inwards.  The total wall thickness of the tube is 109~$\mu$m and the inner diameter is 15.55~mm. The electrical resistance of each straw, from one end to the other, is between 75~$\Omega$ and 100~$\Omega$. The differences are mostly due to variations in the thickness of the aluminum layer.

\begin{figure}[ht!]\centering
  \includegraphics[width=0.45\textwidth]{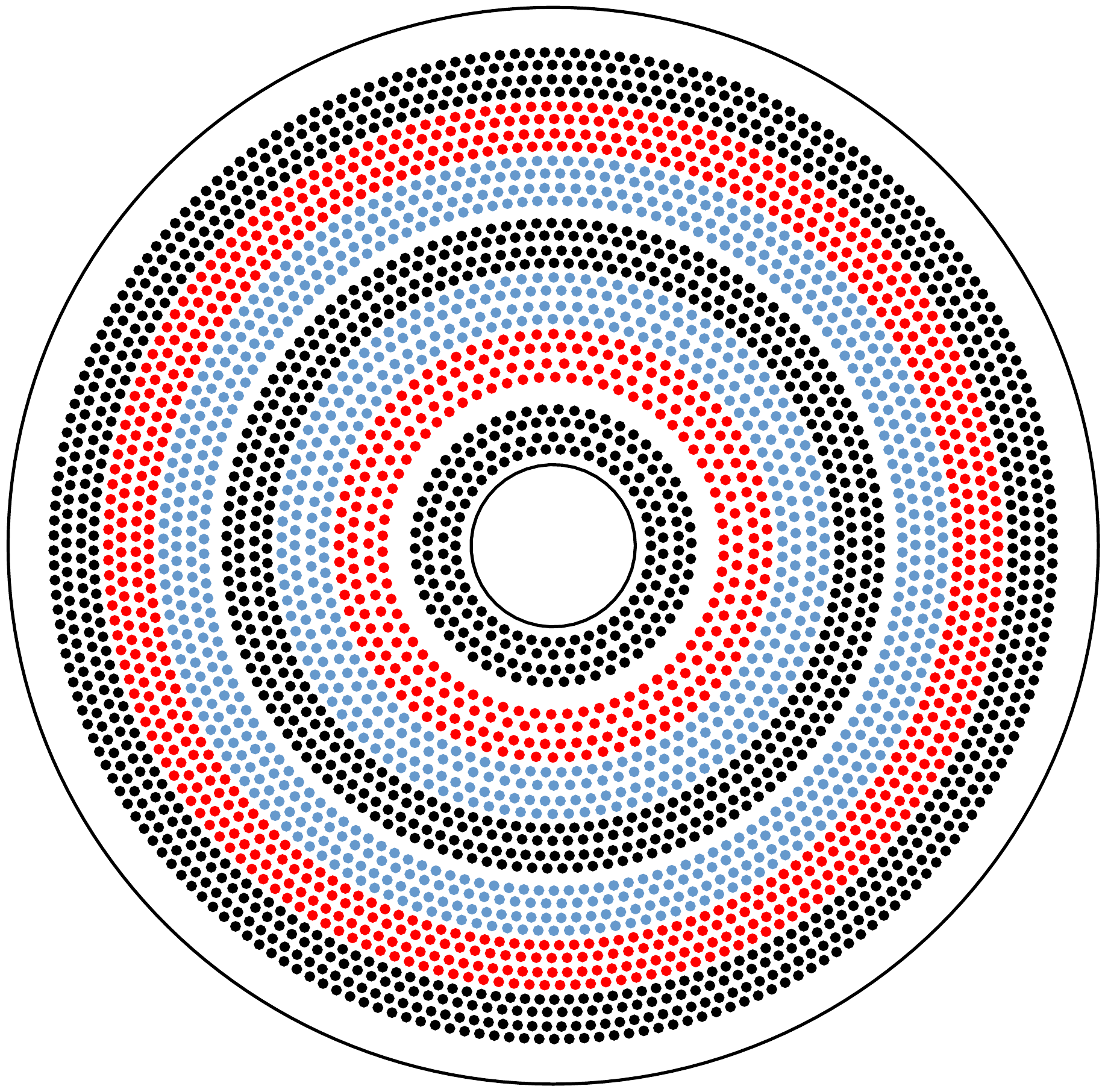}
\caption[]{\label{cdc_strawplac} Diagram showing the position of the straws at the upstream endplate.  The axial straws are shown in black, the $+$6$^{\circ}$ stereo layers are shown in red and the $-$6$^{\circ}$ stereo layers are shown in light blue. } 
\end{figure}

The straws are arranged in 28 radial layers surrounding the inner shell. 12 of the layers are axial (parallel to the beam axis) and the remaining 16 are placed at stereo angles of $\pm$6$^{\circ}$. These are ordered such that the innermost 4 layers are axial, followed by (at increasing radius) 4 layers at $+$6$^{\circ}$, 4 layers at $-$6$^{\circ}$, 4 axial layers, 4 layers at $-$6$^{\circ}$, 4 layers at $+$6$^{\circ}$ and 4 axial layers. This is shown in Fig.~\ref{cdc_strawplac}.  

\begin{figure}[ht!]\centering
\includegraphics[width=0.35\linewidth]{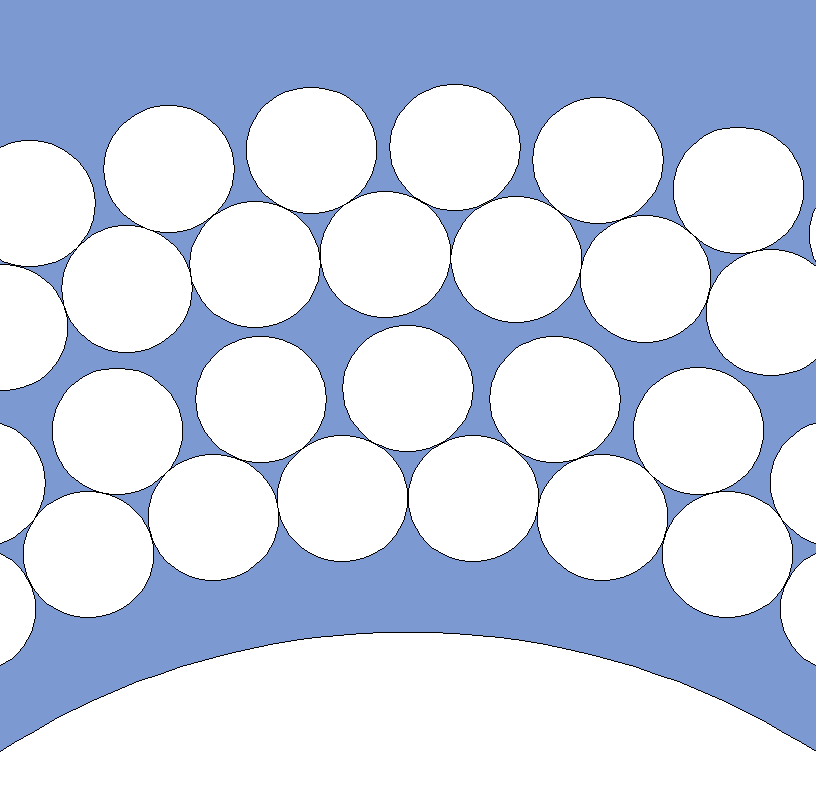}
\caption[]{\label{cdc_holes_sketch} Diagram showing close-packing of the straws in a small section of rows 1 to 4.}
\end{figure}

The layers are paired and located so that the first layer of each pair contains the largest number of straws possible for its radius, and the straws in the second layer are close-packed against those in the first.  This is illustrated in Fig.~\ref{cdc_holes_sketch}. 

\begin{figure}[ht]\centering
\includegraphics[width=0.45\textwidth]{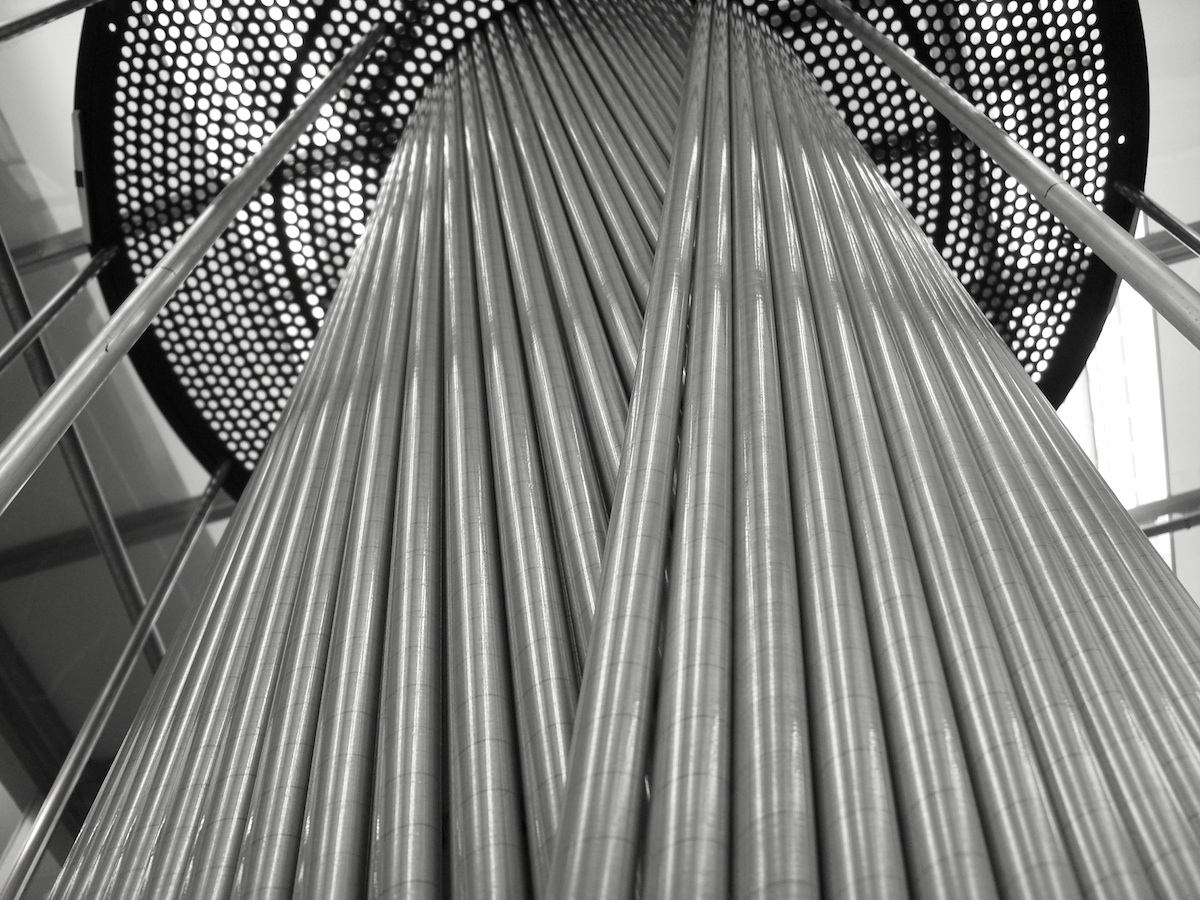}
\caption[]{\label{fig:cdc_photo_stereotubes}Straw-tubes in stereo layers 8 and 9.} 
\end{figure}

\begin{figure}[ht!]\centering
\includegraphics[width=0.45\textwidth]{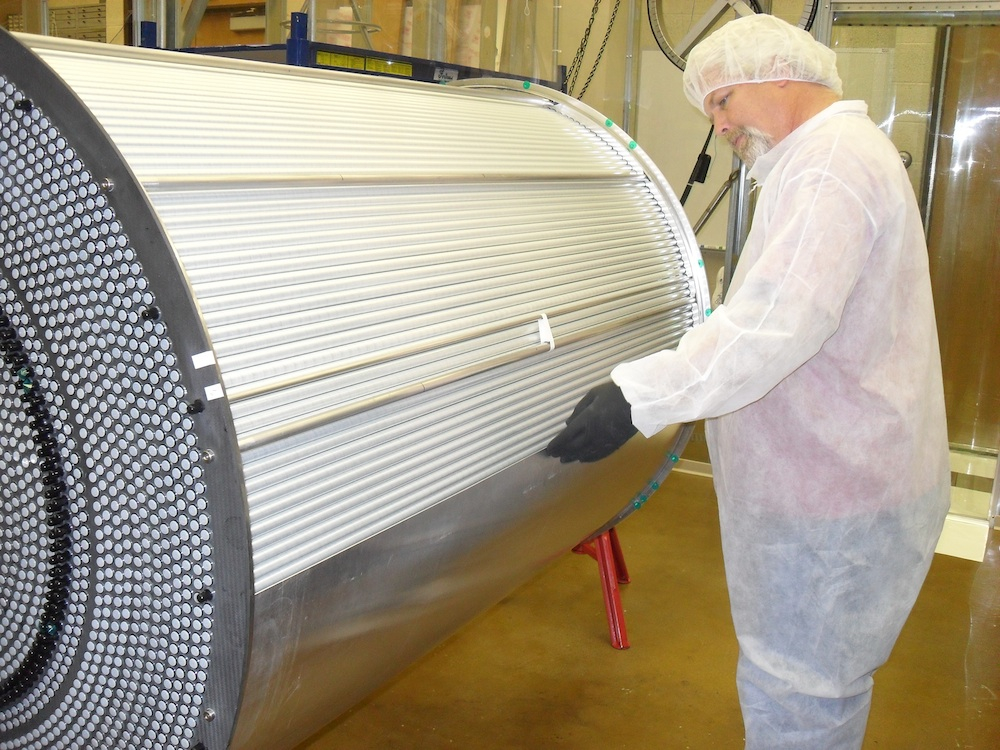}
\caption[]{\label{fig:cdc_photo_outershell}Straw-tubes in layer 28, with one half of the outer shell in place.} 
\end{figure}

Non-conductive epoxy resin\footnote{3M Scotch-Weld DP460NS, www.3m.com} was used to glue each straw-tube to its neighbors within the same layer at three points evenly distributed along its length. In the first layer of each pair, every sixth straw was also glued to the straw behind it. In the second layer of each pair, every straw was glued to the straw behind it. Figure~\ref{fig:cdc_photo_stereotubes} shows straws in opposing stereo layers 8 and 9 and Fig.~\ref{fig:cdc_photo_outershell} shows the outermost row of straws, with the outer shell partly installed. 


\subsection{\label{sec:cdc:straws-wire}Straw and wire assembly}
\begin{figure}[ht!]\centering
\includegraphics[width=0.45\textwidth]{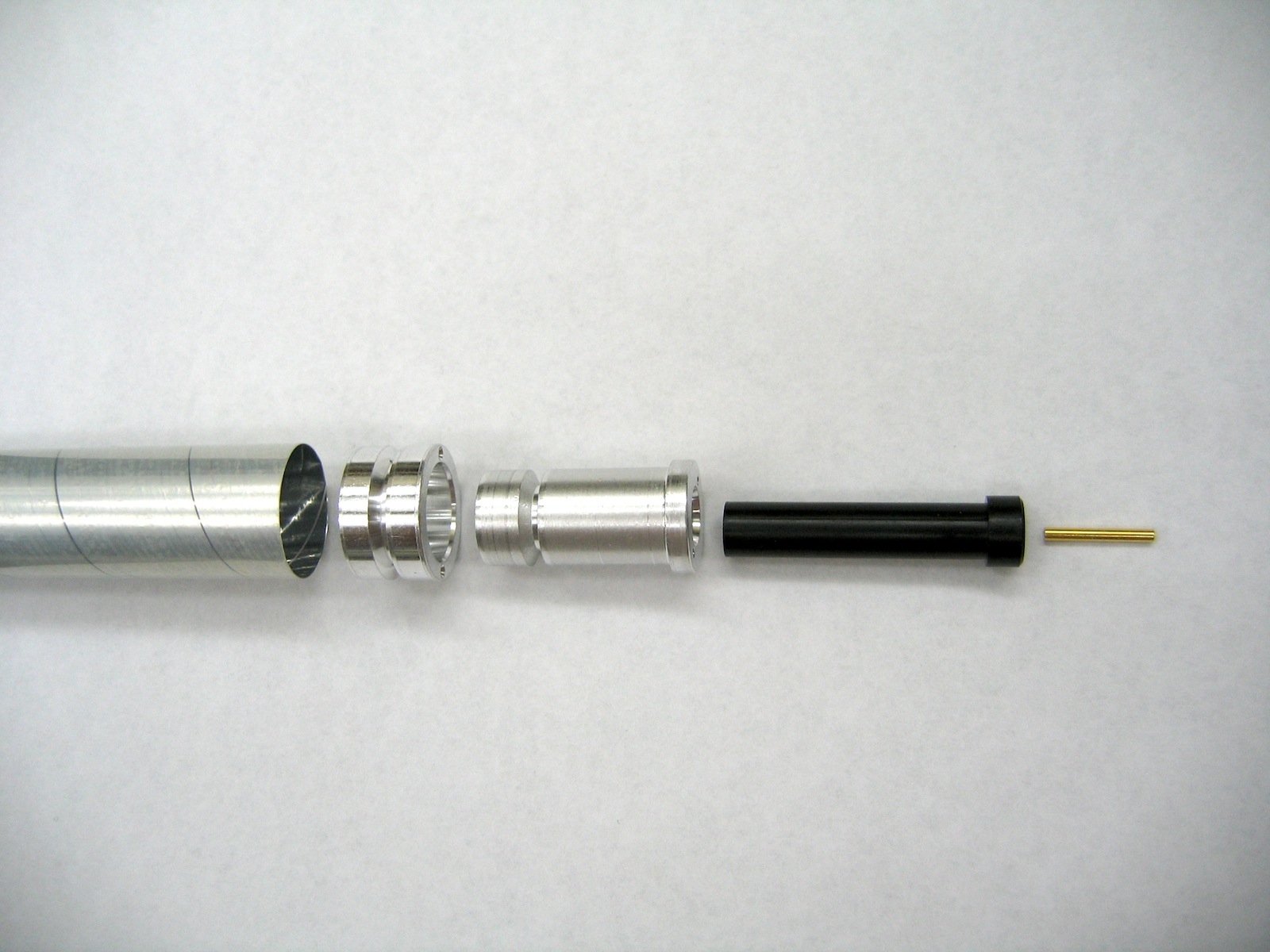}\\
\includegraphics[width=0.45\textwidth]{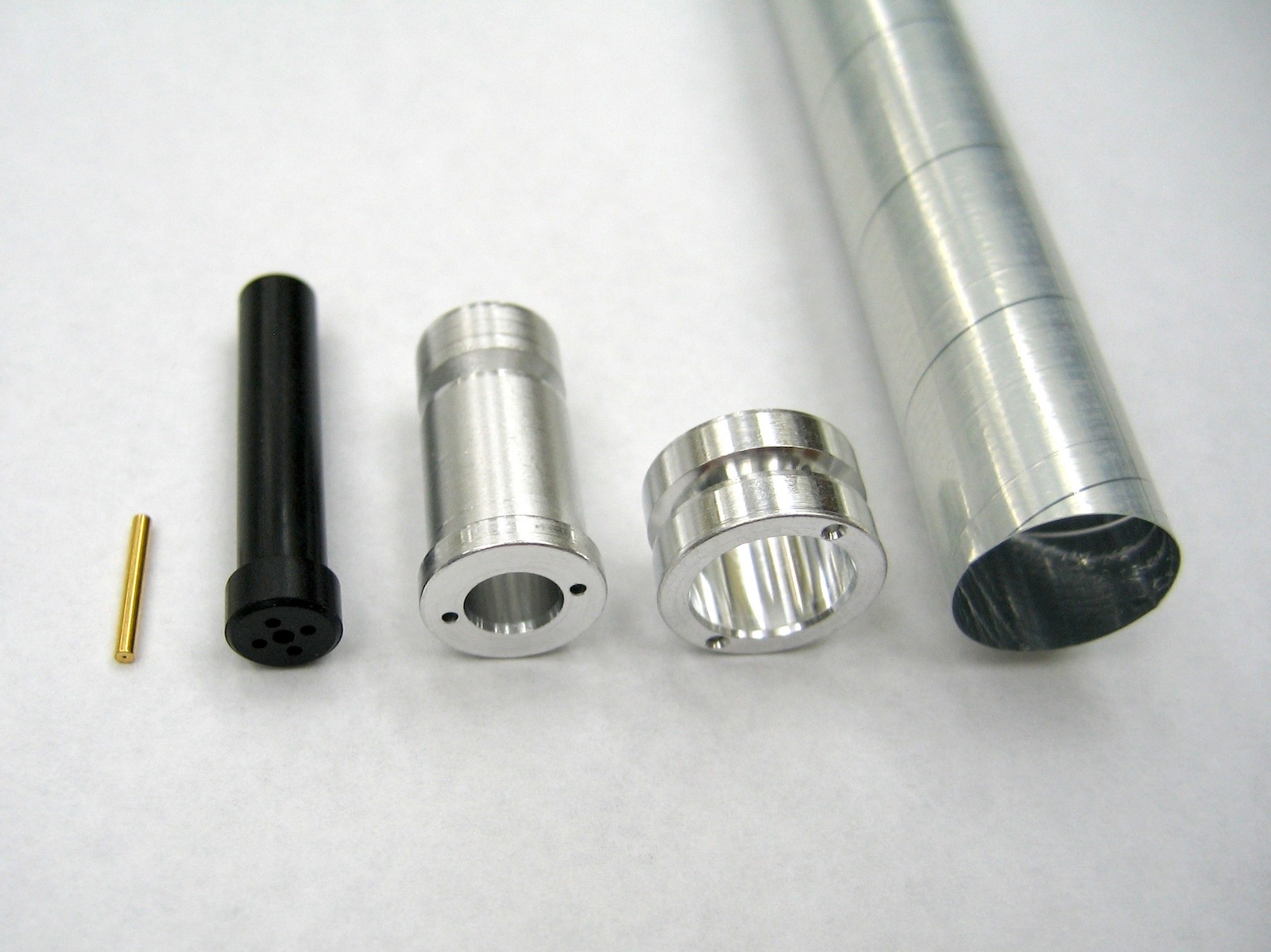}
\caption[]{\label{fig:cdc_photo_strawcpts}Straw, donut, feedthrough, pinholder and crimp pin.} 
\end{figure}

The straw assembly components are shown in Fig.~\ref{fig:cdc_photo_strawcpts}.  A `donut' ring was glued inside each end of the straw. A `feedthrough' tube was glued through the endplate into the donut to hold the straw in position. Two types of donuts, feedthroughs, and epoxy were used: Noryl\footnote{www.sabic.com} plastic donuts and feedthroughs were glued into the carbon fiber endplate with non-conductive epoxy, and aluminum donuts and feedthroughs were glued into the aluminum endplate with silver conductive epoxy\footnote{TIGA 920-H, www.loctite.com}.  Sufficient epoxy was used to make each joint gas-tight.  The conductive epoxy ensures that the electrical grounding of the aluminum endplate is shared with the aluminum feedthroughs, donuts, and the aluminum layer on the inside of the straw.  

The donuts and feedthroughs were manufactured with a recess in their exterior surface, accessible by 2 narrow holes bored lengthwise into the component wall. The holes acted as glue ports, permitting epoxy to be injected into the recess (glue trough) through one hole while air exited through the other.  This enabled the epoxy to create a strong joint by filling the trough completely.  The dimensions of these components are given in Table~\ref{tab:cdc:strawcomponentdimensions}.

\begin{table}[ht]\centering
  \caption[]{\label{tab:cdc:strawcomponentdimensions}Dimensions of the straw assembly components.}
\begin{tabular}{|l|r|r|r|}\hline
  Component & Inner  & Outer  & Length \\
                     & diameter & diameter & \\
                     & mm & mm & mm \\
\hline
Straw (straight)  & 15.55 & 15.77   & 1498.09  \\
Straw (stereo)  & 15.55  & 15.77  & 1505.71  \\
Donut (top)       & 11.11  & 15.75   & 0.63 \\
Donut (rest)      & 11.11  & 15.52  & 8.89 \\
Al feedthrough (top)& 6.35   & 12.70  & 2.54 \\
Al feedthrough (rest)& 6.35  & 11.11 & 21.59 \\
Noryl feedthrough (top)& 6.35   & 12.70  & 2.54 \\
Noryl feedthrough (rest)& 6.35   & 11.11 & 18.03  \\
Pinholder (top) & 1.47   & 7.87   & 3.96  \\
Pinholder (rest) & 5.08  & 6.35   & 29.06 \\
Crimp pin & 0.203  & 1.47  & 12.06  \\
\hline
\end{tabular}
\end{table}

The anode wires are held in place by gold-plated copper crimp pins inside Noryl plastic tubes, `pinholders', inserted into the feedthroughs. The inner diameter of the pinholder is 1.47~mm at the top, making a very close fit with the crimp pin, then after 6~mm of length the diameter is reduced to 1.27~mm for a further 1.63~mm, to support the base of the crimp pin, before opening out to a diameter of 5.08~mm. Each pinholder has 4 additional holes surrounding the crimp pin which permit gas to flow in and out of the straw.

The pins were crimped when the wire was under tension, applied by suspending a 30~g weight from the wire, with the chamber orientated so that the wire was hanging vertically. 20~$\mu$m diameter tungsten wire with a flash coating of gold, supplied by Luma-Metall\footnote{www.luma-metall.se}, was used for the anodes. 

\subsection{\label{sec:cdc:wire-tens}Wire tension measurements}
\begin{figure}[ht]\centering
\includegraphics[width=0.45\textwidth]{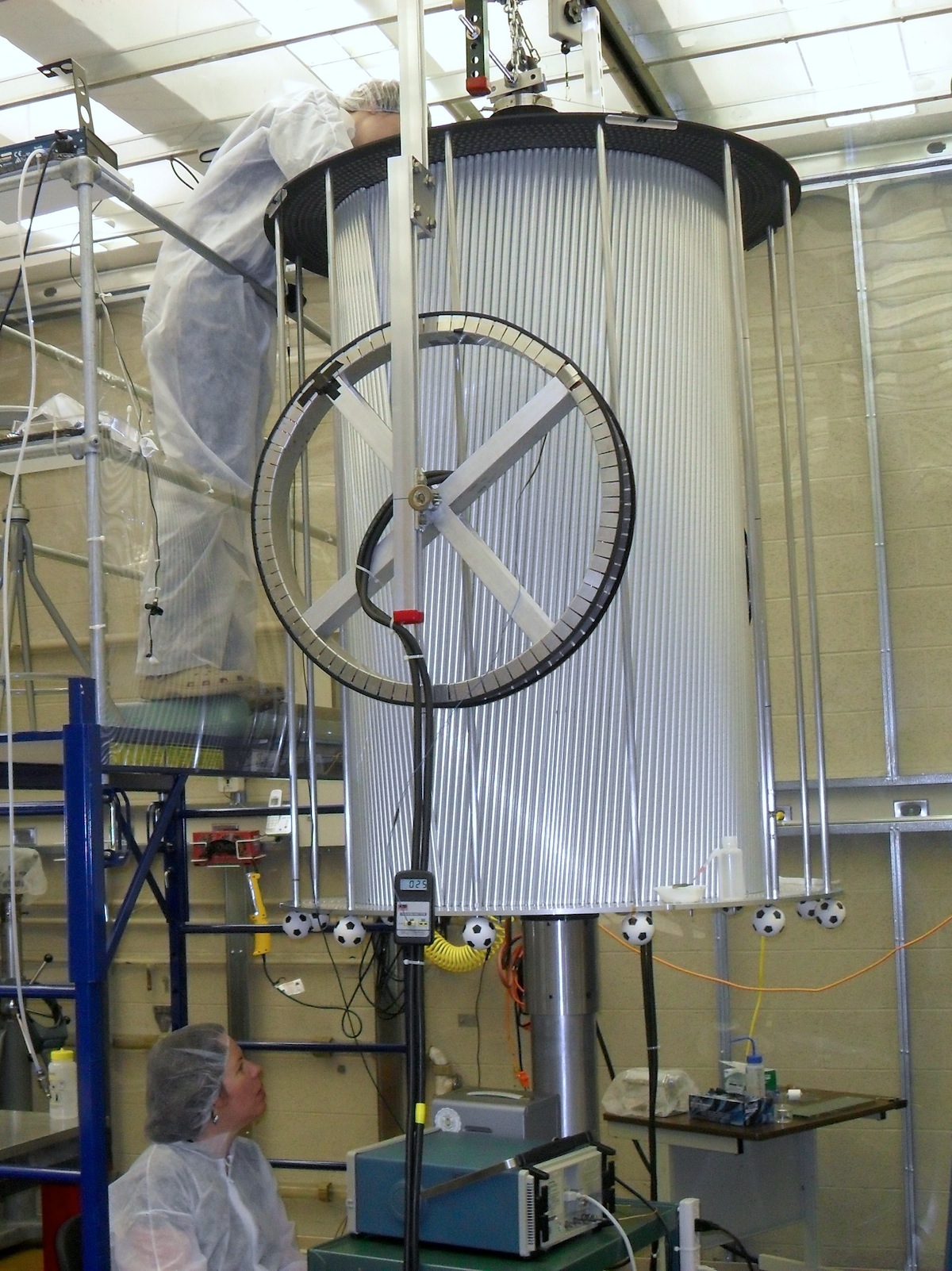}
\caption[]{\label{fig:cdc_photo_tensionmeas}Wire tension measurement being made. One Helmholtz coil is visible in the foreground.} 
\end{figure}

The tension on each wire was measured a few weeks after stringing,  using two Helmholtz coils and a control device which alternated between applying a sinusoidal voltage to the wire and measuring the induced current on the wire.  The wire tension was calculated from the frequency of the applied voltage when the system reached resonance. This technique is described elsewhere~\cite{Roth:1996gn}. The tension measurements were repeated on a monthly basis for 9 months and any wires found to be outside design specifications were replaced. The chamber is shown with the Helmholtz coils in place in Figure~\ref{fig:cdc_photo_tensionmeas}.

A delay in the straw supply during construction led to the innermost 6 axial rows being strung when only 19 of the 28 rows of straws were installed.  The tension measurements were interleaved with the stringing work and after all 28 rows of straws were installed and strung, the 5th and 6th axial rows were found to have low tension.  All the straws in these two rows, any other straws with very low tension and the few straws with broken wires were restrung and remeasured. Most of the wire breakages occurred close to the crimp pin and within a few weeks of stringing.  After that time there were no more breakages for the following year. The tension of each wire was measured at monthly intervals for 8 months and was found to be stable. The range of values measured was 0.265~N to 0.294~N. Two wires broke during the first year of operation at Jefferson Lab but the remaining 3520 were still intact at the time of writing.

\subsection{\label{sec:cdc:gasflow}Gas flow}
Six aluminum tubes of inner diameter 6.35~mm run the length of the CDC, close to the inside wall of the outer cylindrical shell, taking the gas supply from outside the upstream end through the polycarbonate plate and both endplates into the downstream plenum, where the gas enters the straw-tubes through the holes in the pinholders. The gas passes through the straw-tubes into the upstream plenum and then through ten holes in the lower half of the aluminum endplate into the void between the straws and the outer shell of the CDC. Six holes near the top of the aluminum endplate permit the gas to leave the void through exhaust tubes and then bubble through some small containers of mineral oil.  Five thermocouples within each plenum enable the temperature of the gas to be monitored. 
\subsection{\label{sec:cdc:shielding}Electrical shielding}
Electrical shielding was required to minimize the amount of electromagnetic noise picked up by the signal wires. The aluminum endplate provided the common ground for the straw-tubes and also the outer shell, which formed an electrical shield around the tubes. Each half of the outer shell was glued to the aluminum endplate and G-10 outer hub with non-conductive epoxy. In order to ensure a good electrical connection, tabs of aluminum were glued over the joint between the outer shell and the endplate with conductive epoxy at twenty points around the outer radius.

The long straight edges of the two halves of the outer shell were covered with non-conductive glass-cloth electrical tape and then joined together by a 51~mm wide strip of Polyken\footnote{www.polyken.com} 231 tape with a strip of 25~mm wide copper tape glued on top. The copper tape was grounded to the endplate by a tab of copper attached with conductive epoxy. This arrangement ensures that the sidewalls of the cylindrical outer shell have a good connection to the grounded aluminum endplate, while the discontinuity between the two halves of the shell prevents eddy currents from spiraling around the CDC in the event of a magnet quench. For additional reinforcement, a 114~mm wide strip of 0.13~mm thick Kapton\footnote{www.kapton.com} film was glued onto the shell, covering the copper tape along the seam.  The Kapton was glued onto the shell on either side of the tape with DP190\footnote{3M Scotch-Weld DP190, www.3m.com} epoxy.

The upstream outer gas plenum sidewall was covered with 0.13~mm thick copper tape. A copper braid was soldered to the tape at intervals and glued to the aluminum endplate with conductive epoxy. The downstream plenum endwall was constructed from Mylar, aluminized on both sides.  Rectangular tabs extending outwards from the endwall around its radius were glued to the sidewall and outer shell with conductive epoxy.

\begin{figure}[ht]\centering
\includegraphics[width=0.45\textwidth]{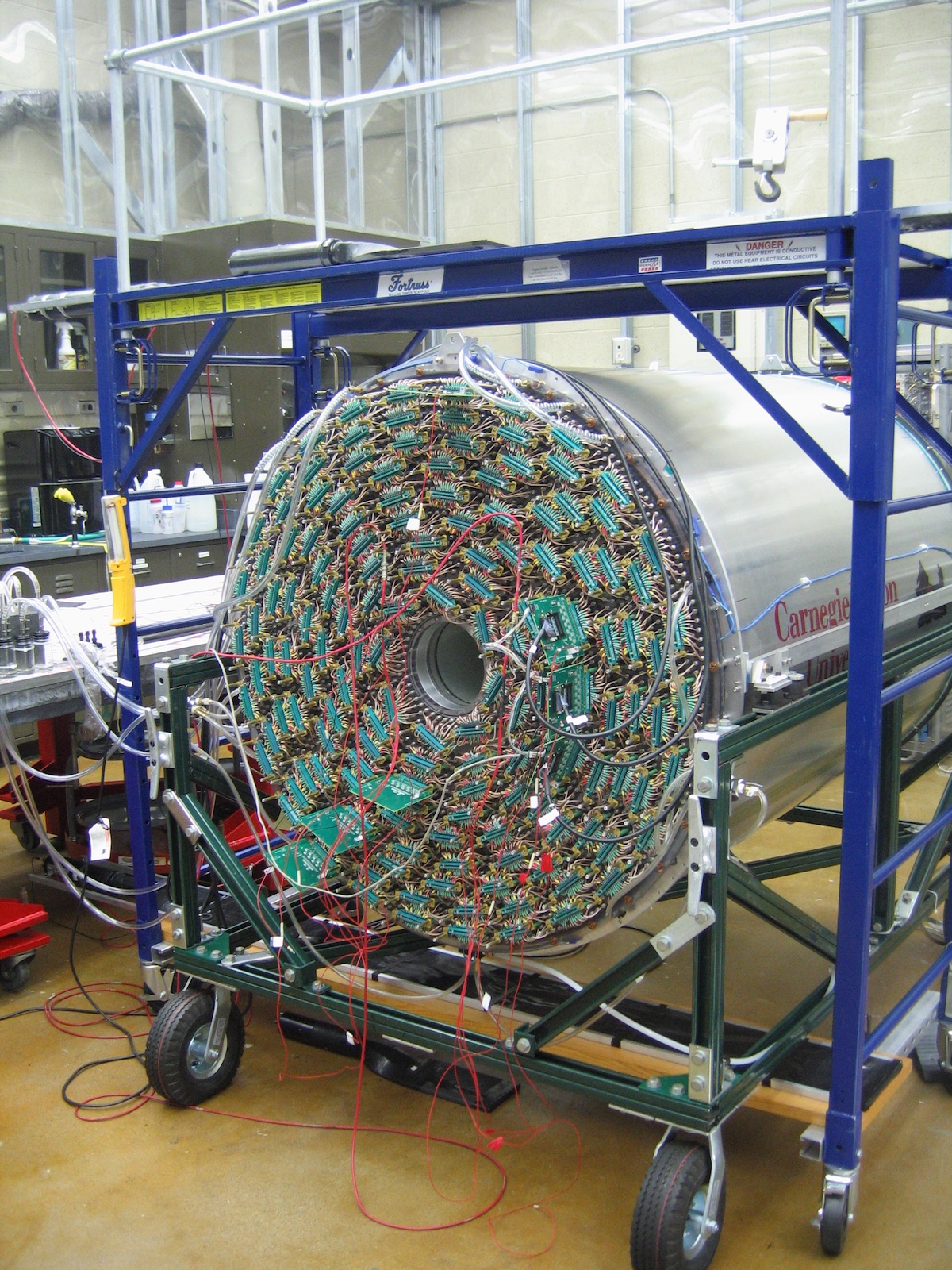}
\caption[]{\label{fig:cdc_photo_oncart}The completed CDC undergoing early readout tests. } 
\end{figure}

Grounded shielded extension cables were used for the downstream thermocouples along the length of the CDC, and for all the thermocouples from the upstream end to the electronics racks, in order to minimize any electrical pickup. Figure~\ref{fig:cdc_photo_oncart} shows the completed CDC undergoing some early readout tests.

\section{\label{sec:cdc:electronics}Electronics}
\begin{figure}[ht]\centering
  \includegraphics[width=0.45\textwidth]{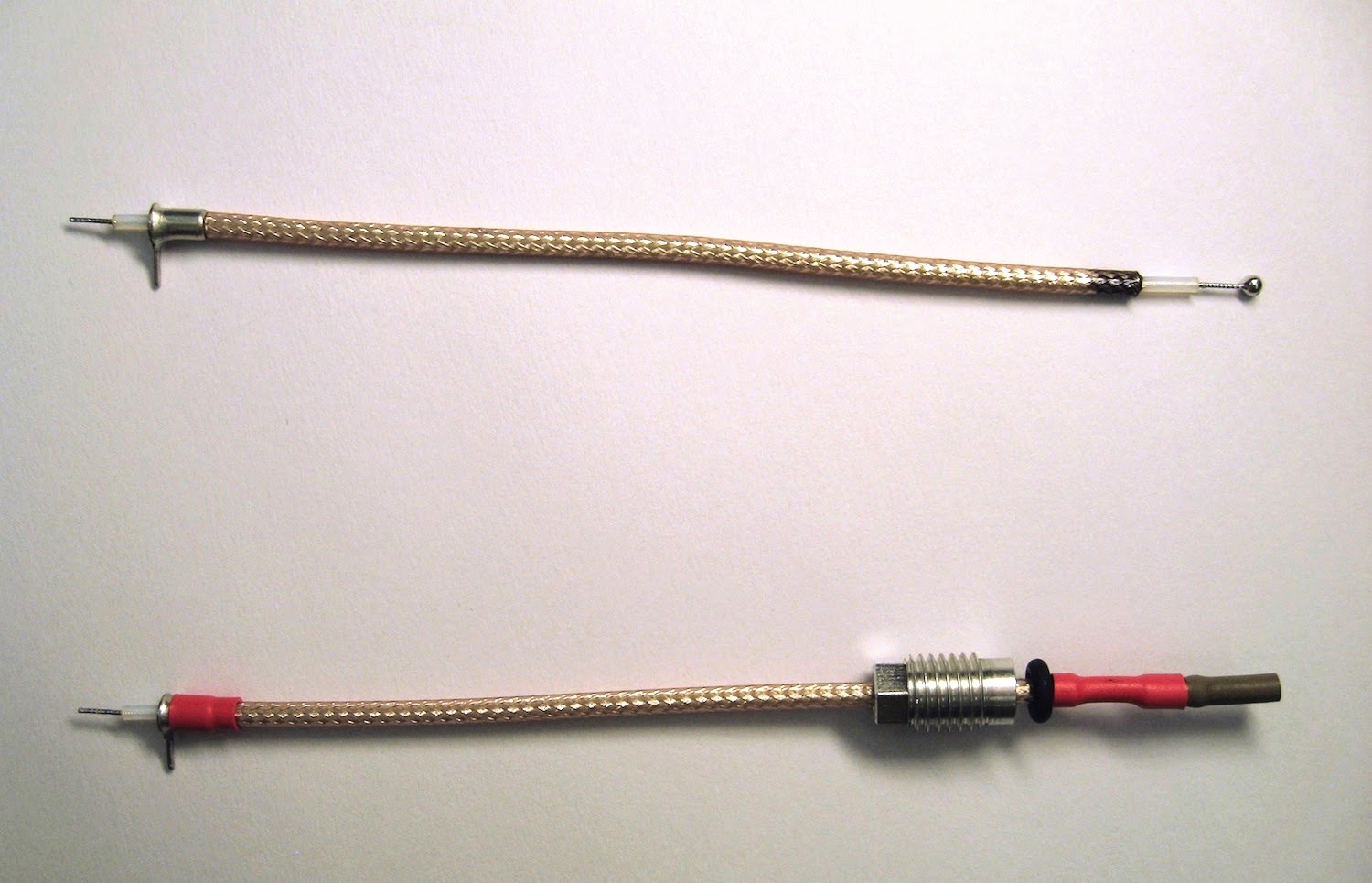}
\caption[]{\label{fig:cdc_photo_hookupwires}Two hookup wires, the upper wire is part-assembled and the silver bead is visible.} 
\end{figure}

The hookup wires, which pass through the polycarbonate endplate and onto the crimp pins inside the upstream gas plenum, were made from RG-316 coaxial cable as follows: at one end of the wire, the inner conductor was exposed for approximately 5~mm and the teflon dielectric was exposed for a further 5~mm.  The end of the shielding braid was sealed with epoxy to prevent gas from migrating along the cable inside the braid. A silver bead was soldered onto the end of the center conductor and then covered with a narrow tube of conductive rubber, approximately 15~mm long, which fits tightly over the bead.  Heat-shrink was then used to seal over the region from the end of the outer covering and braid to the end of the conductive rubber tube.  An O-ring and threaded bushing were fed onto the hookup wire before its other end was finished by stripping back the braid 10~mm and then soldering a ferrule to the braid, then stripping the dielectric 5~mm from the end of the wire.  Two hookup wires, one complete and one partly assembled, are shown in Fig.~\ref{fig:cdc_photo_hookupwires}. The length of wire used for each connection was between 93~mm and 125~mm; this was chosen to be as short as possible, without causing excessive strain on the solder joints. 

The polycarbonate endplate was polished to transparency so that the crimp pins were clearly visible through it.  Each hookup wire was installed by inserting it through a threaded hole in the polycarbonate endplate and sliding the conductive rubber over the corresponding crimp pin until it made a snug fit as the silver bead made contact with the end of the pin. The O-ring and threaded bushing were then fitted into the hole in the endplate, and the ferrule and center conductor at the other end of the hookup wire were soldered onto pads of a transition board.  The transition boards are mounted onto standoffs located on the polycarbonate endplate - one standoff mounts directly to the polycarbonate endplate while the other (grounding) standoff threads onto another standoff which is mounted onto the Al endplate and protrudes through a hole in the polycarbonate endplate, sealed with an O-ring.

\begin{figure}[ht]\centering
\includegraphics[width=0.45\textwidth]{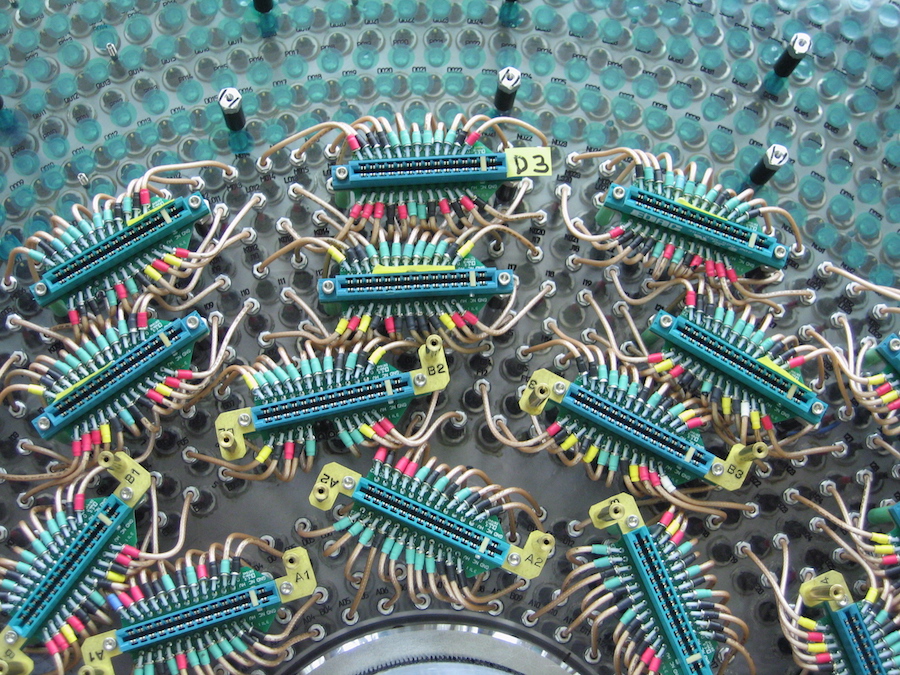}
\caption[]{\label{fig:cdc_photo_transboards}Hookup wires and transition boards.} 
\end{figure}

There are 149 transition boards, each of these was soldered to between 20 and 24 hookup wires, connected to straws from 3 to 4 neighboring rows. Some of the transition boards and standoffs are shown in Fig.~\ref{fig:cdc_photo_transboards}.  Each transition board houses a 30-pin connector for installation of a high voltage board (HVB) that provides approximately 2~kV for up to 24 wires, 2 connections of approximately 2~kV for the shielding braids, 2 ground connections to the grounding standoffs and 2 unused connections which are located between the HV and ground connections. The HVBs also house the preamplifier cards~\cite{Barbosa:gluexdoc2515}.  

The preamplifiers have 24 channels per board, are charge-sensitive, and are  capacitively coupled to the CDC. The preamplifiers are connected to 125 MHz flash analog to digital converters (fADC)~\cite{Visser:2010zpa}, with three preamplifiers to each fADC. The high voltage (HV) supply units used are CAEN A1550P\footnote{www.caen.it} with noise-reducing modules added to each crate chassis. The low voltage (LV) supplies are Wiener MPOD MPV8008\footnote{www.wiener-d.com}. 

\section{\label{sec:cdc:fadc}fADC readout and timing}
The fADCs use Xilinx\footnote{www.xilinx.com} Spartan-6 (XC6SLX25) FPGAs for signal digitization and on-board data processing at 125\,MSPS with 12 bit resolution. Following each trigger, if a pulse is found, the readout data for each channel contain the pulse time, pulse amplitude, pulse integral, pedestal before the start of the pulse, a quality code for the time measurement and the number of overflow samples within the integral period, fitted into two 32-bit words. The pulse time is measured with 0.8~ns precision. The fADCs can also output the data in a much longer diagnostic format, which appends the raw sample data to its usual compact output.  

Configuration parameters allow the user to specify the start of the pulse data window relative to the trigger arrival time, the length of the window, the number of samples used for the pedestal calculation, the number of samples by which the local pedestal calculation precedes the pulse threshold crossing signal, the factors by which the pulse amplitude, integral and pedestal are scaled down before output by right-shifting, and the thresholds that are used for pulse identification and timing.

The trigger signal prompts the firmware to search through the data window for a pulse, which is found if two or more consecutive samples exceed the pedestal at the start of the data window by the hit threshold value or more.  If a pulse is found, the local pedestal is calculated a few samples (configurable) before the hit threshold crossing sample. The samples immediately after this are searched again to find the start of the pulse, first by searching forward to find the first sample value that exceeds the local pedestal by a high timing threshold or more, and then back from this point to find where the signal value exceeds the local pedestal by low timing threshold or less. Looking for the larger threshold crossing and backtracking to find the lower threshold crossing ensures that the edge of the pulse has been found and not a smaller fluctuation. The first sample that is at or above the low timing threshold crossing is upsampled by a factor of 5 and then interpolated to find the threshold crossing point in units of sample/10 (0.8~ns).  Various quality checks are made throughout the pulse analysis and if any are failed then the quality code bit is set.  In certain cases, such as sample values of zero, or pedestal values above a set limit, the firmware does not look for the time threshold crossings but returns a time value of the hit threshold crossing time minus a constant to indicate the error condition.  Similarly, if any problems are encountered during upsampling (for example, unrealistic steps in signal value that could be caused by faulty connectors) then the time value returned contains the low time threshold sample and a code indicating the error condition.
  
Signal integration starts with the low timing threshold crossing sample and ends at the end of the pulse data window. If any overflow samples are found during the integration then the overflow count is incremented, up to a maximum of 7.  If the pulse height, integral or pedestal values are too large to fit into their allocated space in the output words, their output has all bits set.

In the analysis software, the pulse time quantity returned by the algorithm is converted to a time measurement by multiplying by 0.8~ns, which is one tenth of the sample period. It includes a constant offset, corresponding to the earliest possible drift time.  This offset is determined during offline analysis and subtracted from the drift time returned by the algorithm to give the net drift time.

\section{Operating Conditions}
\subsection{Gas system}
The gas used is a mixture of 50\,\%~CO$_{2}$ and 50\,\%~Ar at atmospheric pressure. A small amount (approximately 0.5\,\%) of isopropanol is added by bubbling the gas through a bath of isopropanol at 40F to prevent loss of performance due to aging\cite{KADYK1991436,VAVRA20031}. 
The gas flow rate is 1.3 volumes/day, ie. the gas in the entire active volume of the chamber is replaced with a frequency of 1.3/day. 
\subsection{Operating voltage}
\begin{figure}[ht]\centering
\includegraphics[width=0.45\textwidth]{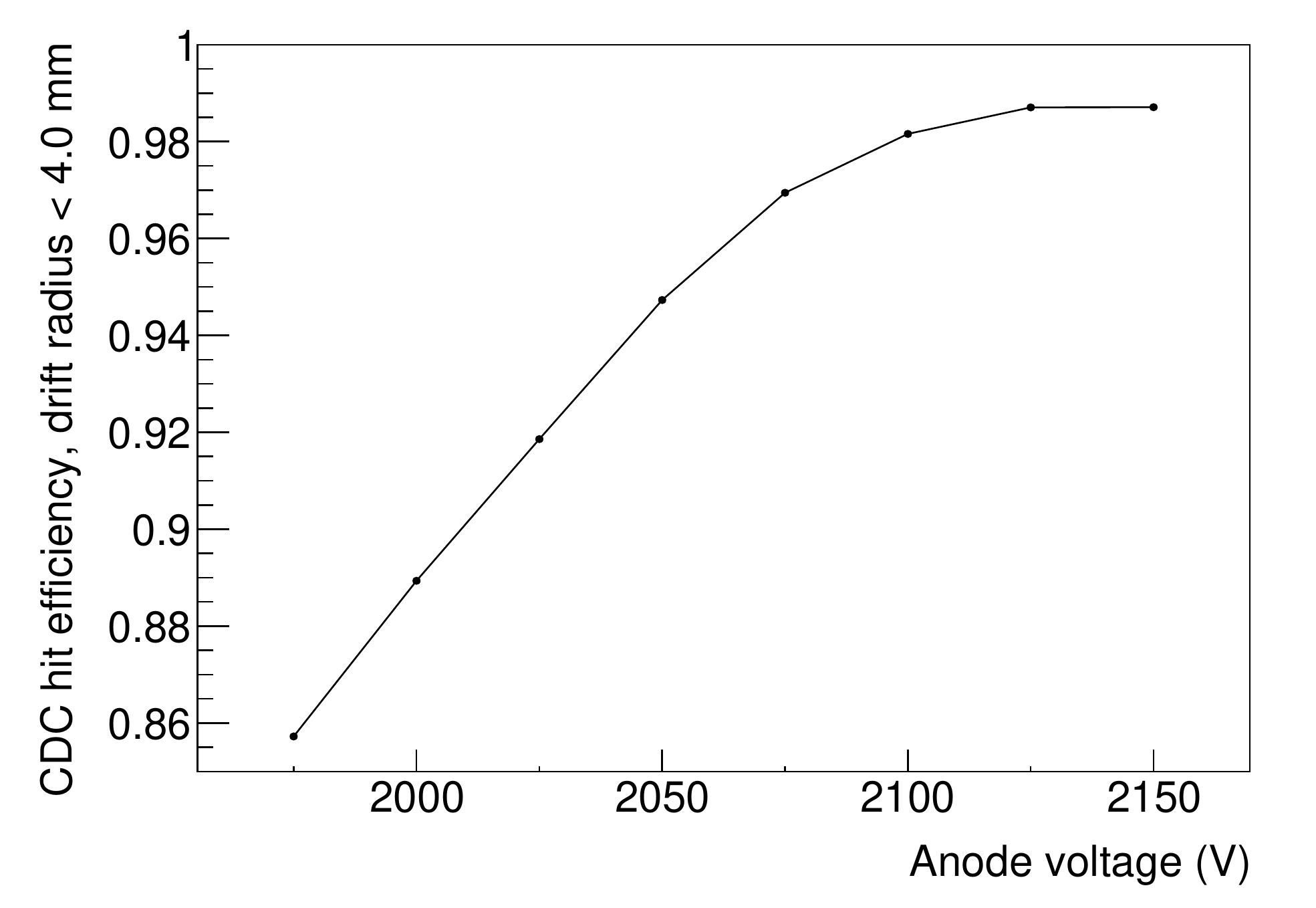}
\caption[]{\label{fig:hv_scan}Hit efficiency for tracks within 4~mm of the wire, as a function of anode voltage. The operating voltage chosen is 2125~V.} 
\end{figure}

The operating voltage on the anode wires was determined by performing a high voltage scan and examining the hits-per-track efficiency, which is the ratio of found to expected hits in the straws along a track, at each voltage. As the voltage increases, at first the efficiency increases dramatically, and then it reaches a plateau and remains stable for a while. This is illustrated by Fig.~\ref{fig:hv_scan} which shows hit efficiency as a function of high voltage. The voltage at the start of the plateau was chosen to be the operating voltage; this is 2125~V. 

The current on each HVB is limited to 10~$\mu$A. Under normal running conditions the currents are comfortably below this level, with 5~$\mu$A on the HVBs (24 channels each) closest to the beam. 

\subsection{Cross-talk}
There is no cross-talk between neighboring straw-tubes. However, following a very saturated signal on one channel, cross-talk can occur from 16 to 40~ns later in the form of small near-simultaneous signals on the other channels connected to the same preamplifier, as they share common HV and ground lines. These secondary pulses are identified and eliminated by the analysis software.

\subsection{Drift Velocity}
Due to the inverse-radius dependence of the electric field in the straw-tubes, the drift velocity of the electrons is not constant. In the case of no magnetic field, the average over the central region of the straws is $15$~mm/$\mu$s, with values rising rapidly very close to the anode wires at the center of the straw. In normal chamber operation with the magnetic field on, the drift velocity is about $95$\% of the zero-field value and the Lorentz angle is typically $14^{\circ}$. Thus, with field on, the electrons have both a slower drift velocity and follow a curved path to the anode, leading to longer drift times from the same ionisation point in the straw. In going from no magnetic field to nominal magnetic field, the maximum drift time increases between $30$ and $55$~ns, depending on the local magnetic field strength.

\subsection{Environmental and Control System Data Monitoring}
An archive is maintained of data from sensors in the control system, including gas flow rates, high and low voltages and total current drawn by all of the anode wires connected to each high voltage board. The archive also includes the temperature measurements from the thermocouples in the CDC's gas plenums and the atmospheric pressure. The temperature in the experimental hall is controlled, the CDC electronics are cooled by forced air flow and the temperature readings from the thermocouples are stable.

The environmental data are also recorded in our physics datastream at 5~minute intervals. Recording the environmental data in the datastream gives us the capability to calibrate the detector gain within each run using the relationship between gain and pressure/temperature but to date this has not been necessary as a time limit of 2 hours is imposed on the data-taking runs and one calibration for each run has been sufficient. The relationship is used for initial gain calibrations.

\section{Calibration and Alignment}
\subsection{Time to Distance Calibrations}
The cylindrical symmetry of a straw-tube with a central wire leads to circular isochrones, where all ionization electrons from a circle arrive at the wire at the same time. Thus, the most basic calibration is a time-to-distance relationship, $d_{0}(t)$. Time-to-distance functions are obtained using tables of values generated from Garfield~\cite{garfield} simulations for typical pressure and temperature that provide a good starting point for calibrations.

\begin{figure}[ht]\centering
\includegraphics[width=0.45\textwidth]{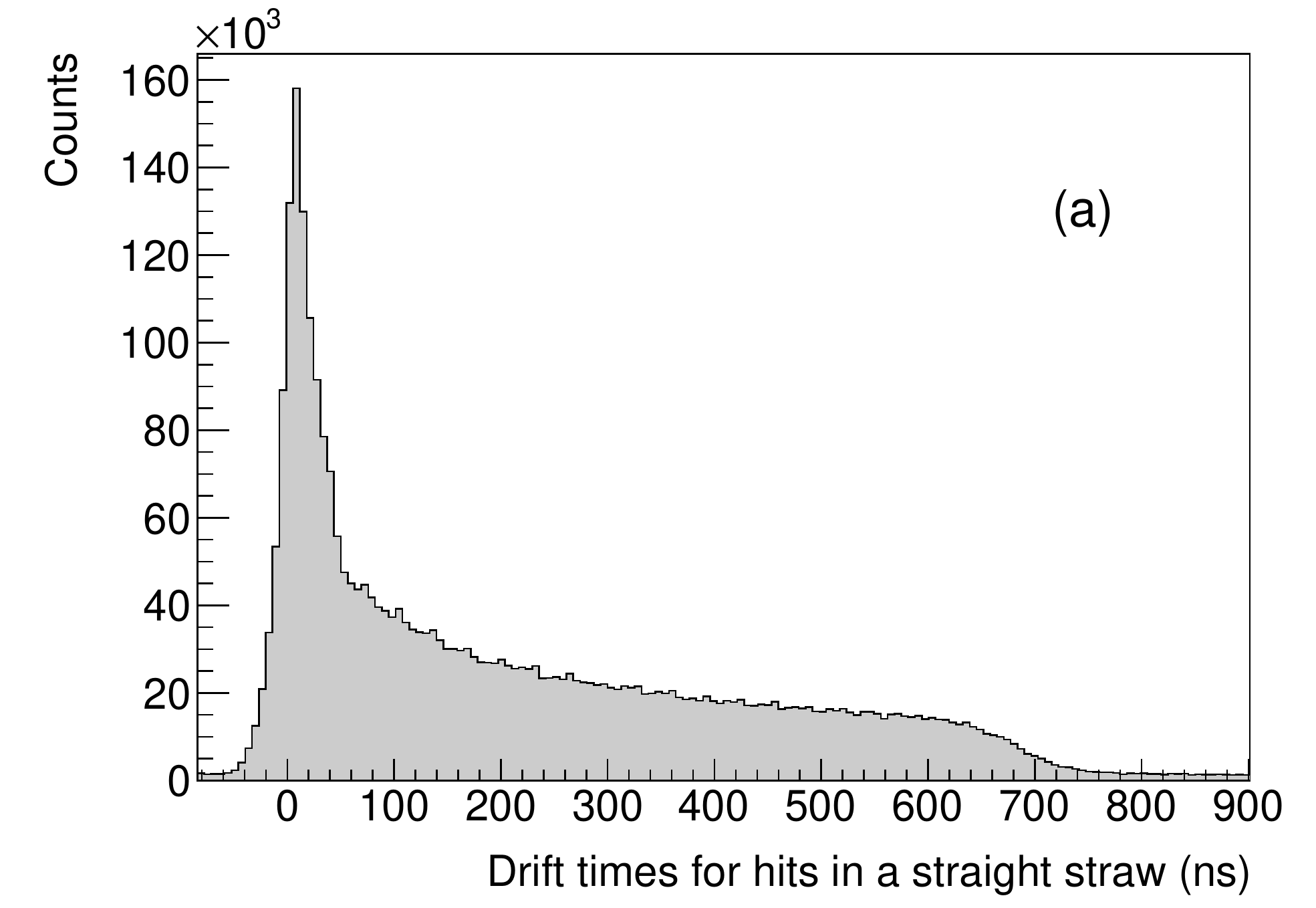}\\
\includegraphics[width=0.45\textwidth]{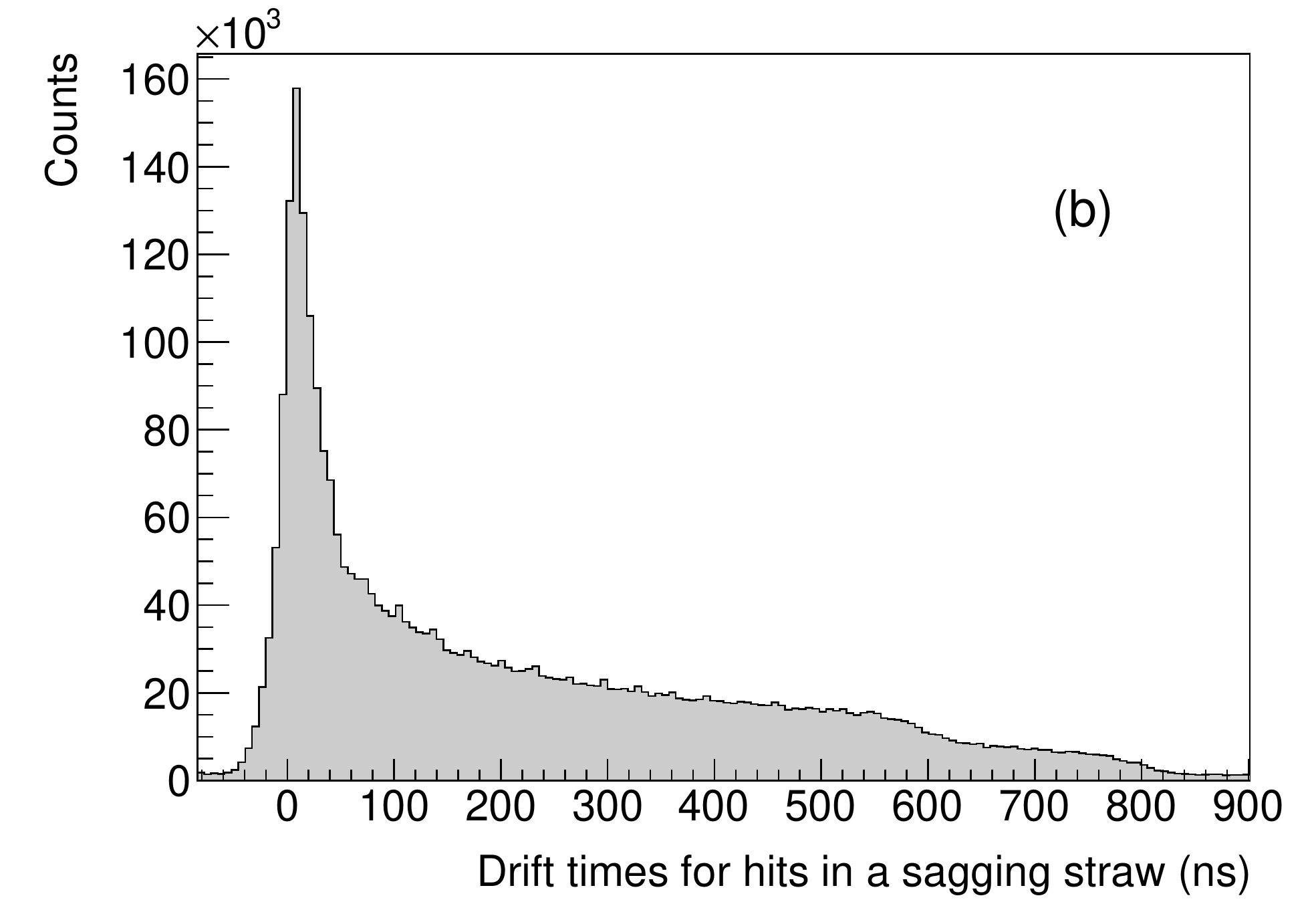}
\caption{Drift time histograms for two nearby straws in the drift chamber (a) one straight and cylindrical and (b) one sagging.}
\label{fig:drift_times}
\end{figure}

During early tests of the CDC, significant variation was found in the maximum drift times between the straws, with many straws showing drift times that were much larger than expected. 
After the CDC had been relocated to Jefferson Lab and installed in Hall D, a large quantity of cosmic ray data were collected from the entire chamber, using the barrel calorimeter as a trigger. The drift time distributions from some of the straws showed good agreement with those predicted using Garfield~\cite{garfield} and could be described using the function given in reference \cite{Avolio:2004dz}, with a maximum drift time of approximately 700~ns. Drift times from one such straw are shown in Fig.~\ref{fig:drift_times}(a). The majority of the straws showed drift time distributions with a much more gradual fall-off at large times, and some of them showed distributions in which an intermediate step down could be seen before the final slope down towards the maximum drift time. An example of this is shown in Fig.~\ref{fig:drift_times}(b).
This effect was caused by straws that were not perfectly cylindrical, with some displacement, or sagging, midway along their length. 

\begin{figure}[ht]\centering
\includegraphics[width=0.45\textwidth]{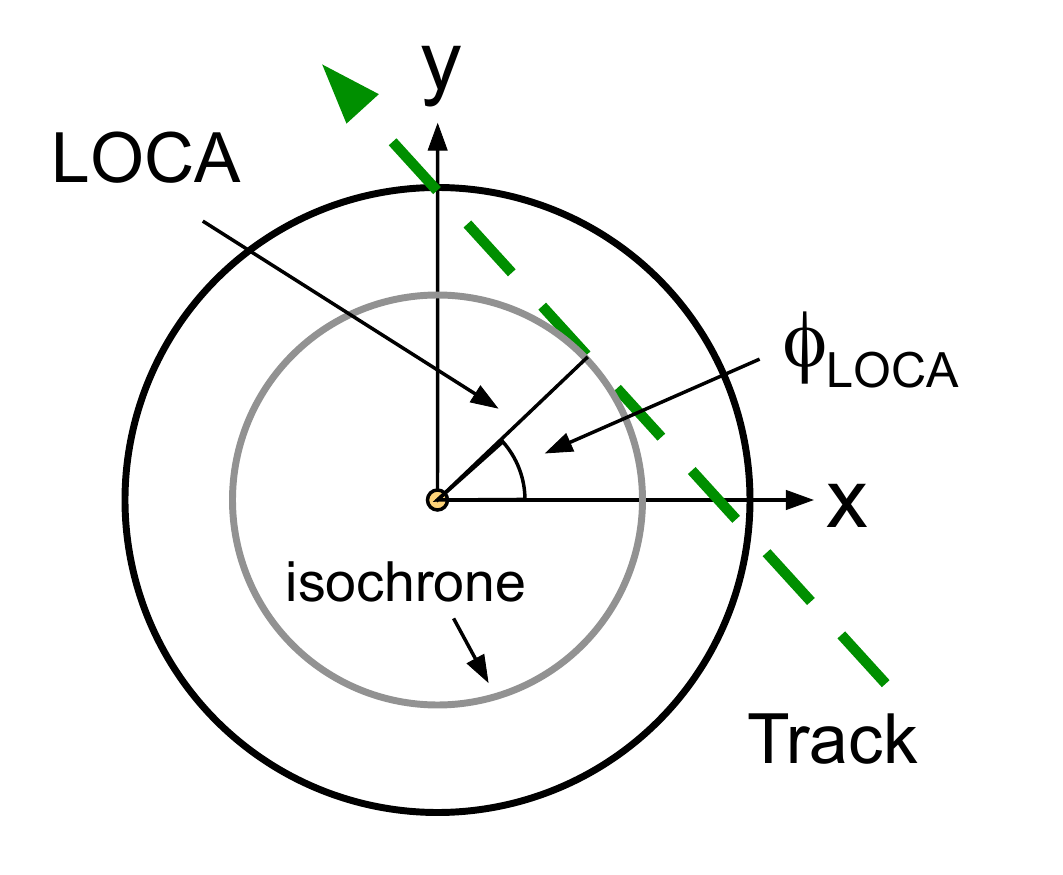}
\caption[]{\label{fig:loca}Cross-section of a straw-tube with the anode wire in the center. A track is shown passing
  through the tube, tangential to a circular isochrone. The line of closest approach, LOCA, is the shortest path from the track to the sense wire. The angle $\phi_{LOCA}$ orients the LOCA.}
\end{figure}

\begin{figure}[ht]\centering
\includegraphics[width=0.45\textwidth]{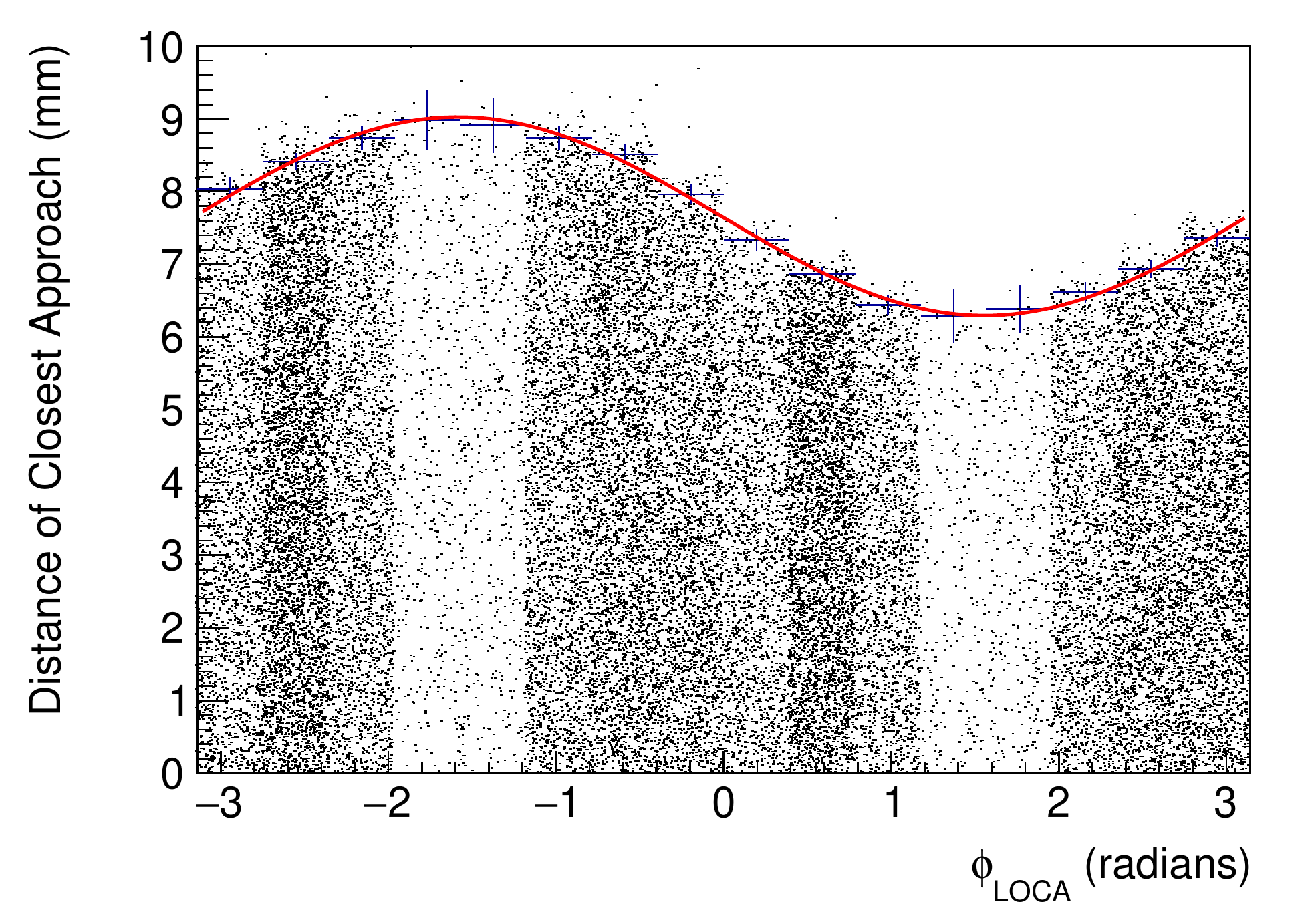}
\caption{The azimuthal dependence of the maximum drift distance for a single straw, shown by a scatter-plot of drift distance obtained from the tracking software vs $\phi_{LOCA}$. The fit function obtained for this straw is $7.66 + 1.37 cos(\phi_{LOCA}+ 1.59)$.} 
\label{fig:loca fit}
\end{figure}

The following procedure was developed to determine the direction and magnitude of the deflection for each straw and was carried out soon after installation. Since the ends of the straws are fixed at the endplates, the bowing of the straws could be modeled as
\begin{equation}
\delta(z,\phi_{LOCA})=c_1\left(1-\left(\frac{z-z_{center}}{L/2}\right)^2\right)\cos(\phi_{LOCA}+c_2),
\end{equation}
where $z_{center}$ is the z-position of the center of the CDC, $L$ is the distance between the endplates, and $\phi_{LOCA}$ is the azimuthal angle of the line of closest approach (LOCA) of a track passing through the straw with respect to the wire position. Figure~\ref{fig:loca} shows the LOCA for an ideal straw. For a bowed straw, $\delta(z,\phi_{LOCA})$ is the change in length of the LOCA caused by the straw deformation, which is specific to the z-position and the angle of the LOCA. 
The constants $c_1$ and $c_2$ were determined for each straw by fitting straight tracks from cosmic events with the solenoid off and filling a histogram of the drift distance calculated for the fitted track as a function of $\phi_{LOCA}$. The values of $c_1$ and $c_2$ were found by fitting the edges of the distribution to a function of the form
\begin{equation}
f(\phi_{LOCA})=c_0+c_1\cos(\phi_{LOCA}+c_2).  
\end{equation}
A sample fit is shown in Fig.~\ref{fig:loca fit}.  

\begin{figure}[ht]\centering
\begin{tabular}{c}
\includegraphics[width=0.45\textwidth]{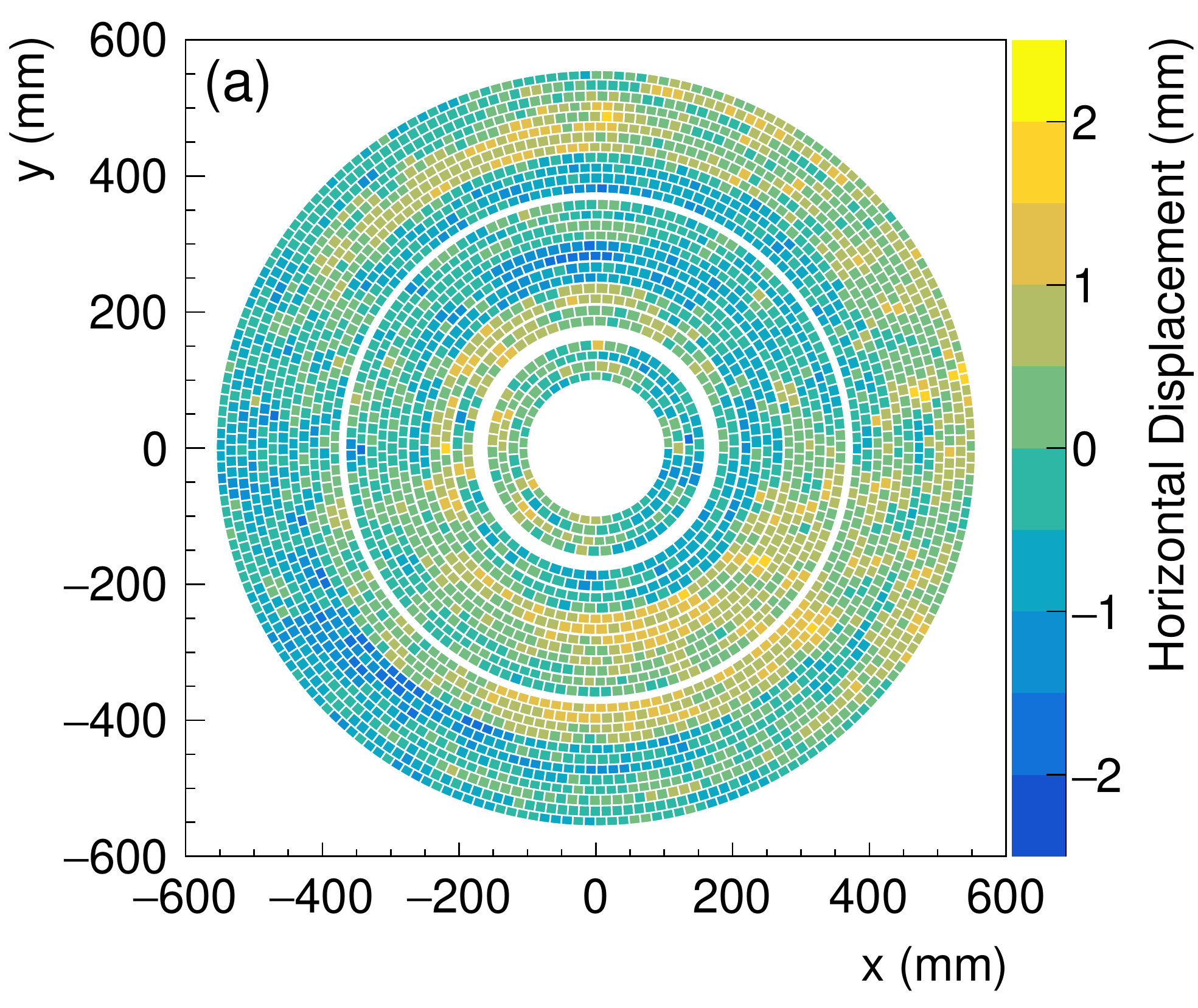}\\
\includegraphics[width=0.45\textwidth]{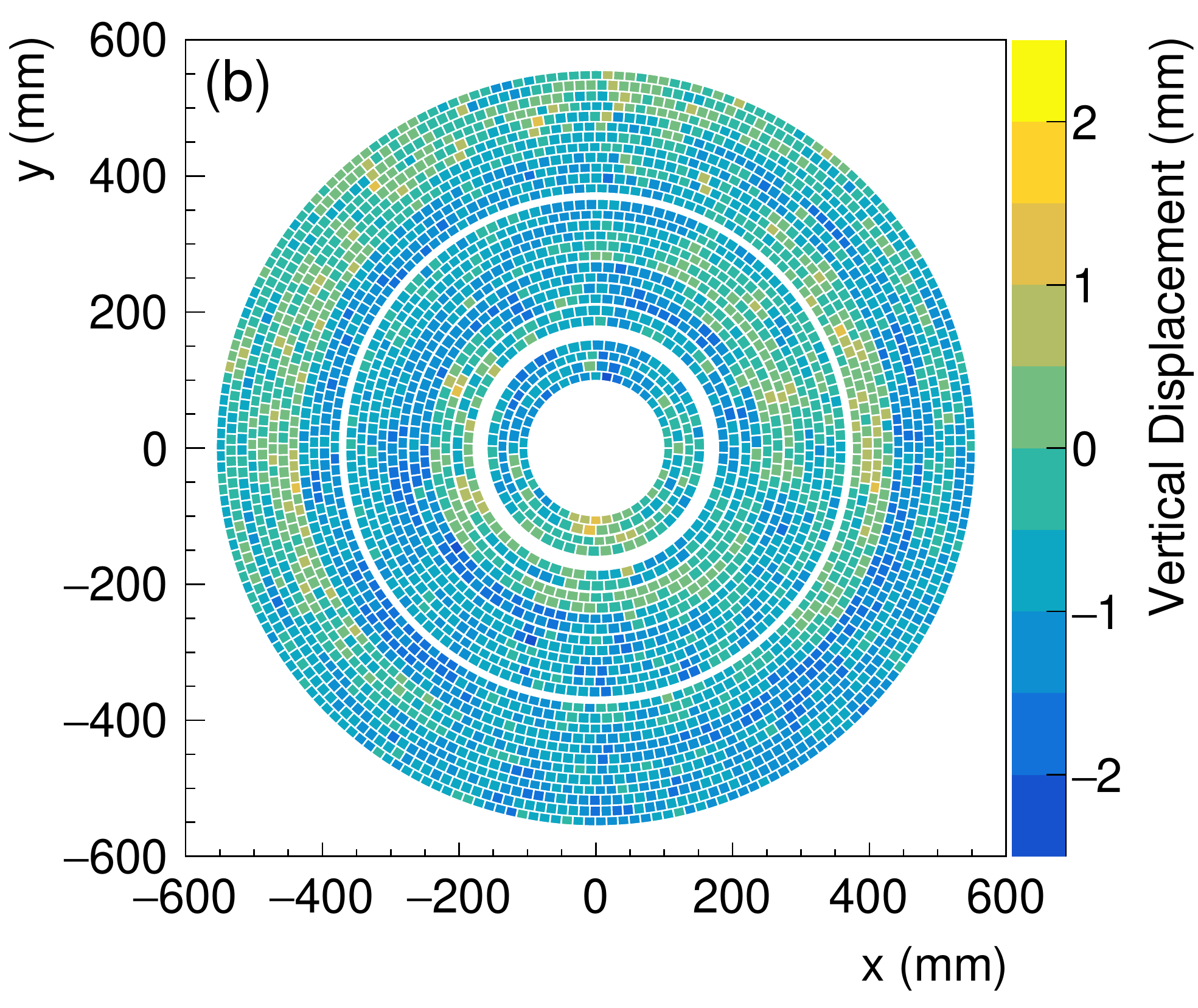}
\end{tabular}
\caption{Schematic of the CDC, looking upstream, using the color scale to show the maximum deflection of each straw in the (a) horizontal and (b) vertical directions.}
\label{fig:delta projections}
\end{figure}

Figure~\ref{fig:delta projections} shows the maximum displacement of each straw in the vertical and horizontal directions; the displacement is mostly in the vertical direction, and it is more common in the stereo layers than the straight layers. The mean displacement was found to be approximately 1~mm, with a range from 0 to 2.1~mm. The procedure to determine the straw sag was repeated a year after the first evaluation and no significant difference was found between the two studies.  

The following section describes how the measurements of straw sag were used to develop a function relating the drift distance to the drift time.  The straw sag has a significant impact on the drift time because the wire is no longer necessarily at the center of the straw.  In general, the time-to-distance relationship depends on both $z$ and $\phi_{LOCA}$. A function of the form
\begin{equation}
d(t)=f_\delta\left(\frac{d_0(t)}{f_0}P+1-P\right),    
\end{equation}
has been found to describe this accurately. In this formula, $d_0(t)$ is interpolated from a table of time-to-distance for an ideal straw, and 
\begin{equation}
P=\left\{\begin{array}{cl}
0 & t>T \\
\frac{T-t}{T} & t\le T 
\end{array}
\right.
\end{equation}
with $T=$250~ns.  250~ns was chosen as the transition point because drift times less than this are not affected significantly by the distortion of the electric field due to the straw sag. It corresponds to a drift distance of approximately 5~mm.  If $\delta(z,\phi_{LOCA})\ge 0$, the distance to the wire is larger than the ideal case.  For this case,
\begin{eqnarray}
f_\delta&=&a\sqrt{t} + bt + ct^3,\\
f_0&=&a_1\sqrt{t} + b_1 t + c_1 t^3,\\
a&=&a_1 + a_2 |\delta|,\\
b&=& b_1 + b_2 |\delta|,\\
c&=& c_1 + c_2 |\delta| + c_3 \delta^2.
\end{eqnarray}
If $\delta<0$, we use
\begin{eqnarray}
f_\delta&=&a\sqrt{t} + bt, \\
f_0 &=& a_1\sqrt{t} + b_1 t, \\
a &=& a_1 + a_2 |\delta| + a_3 \delta^2, \\
b &=& b_1 + b_2 |\delta| + b_3 \delta^2 .
\end{eqnarray}

\begin{figure}[ht]\centering
\includegraphics[width=0.45\textwidth]{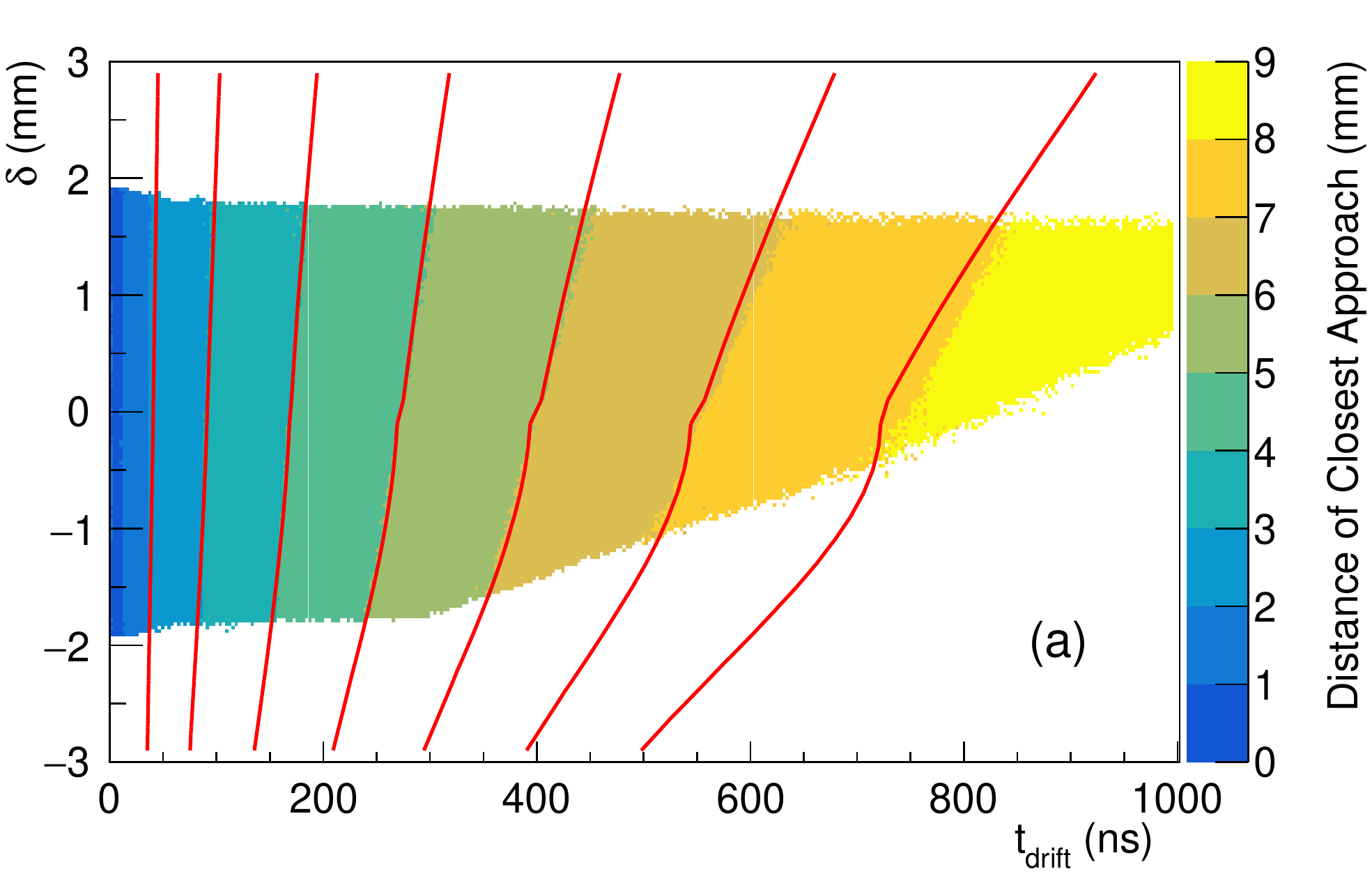}\\
\includegraphics[width=0.45\textwidth]{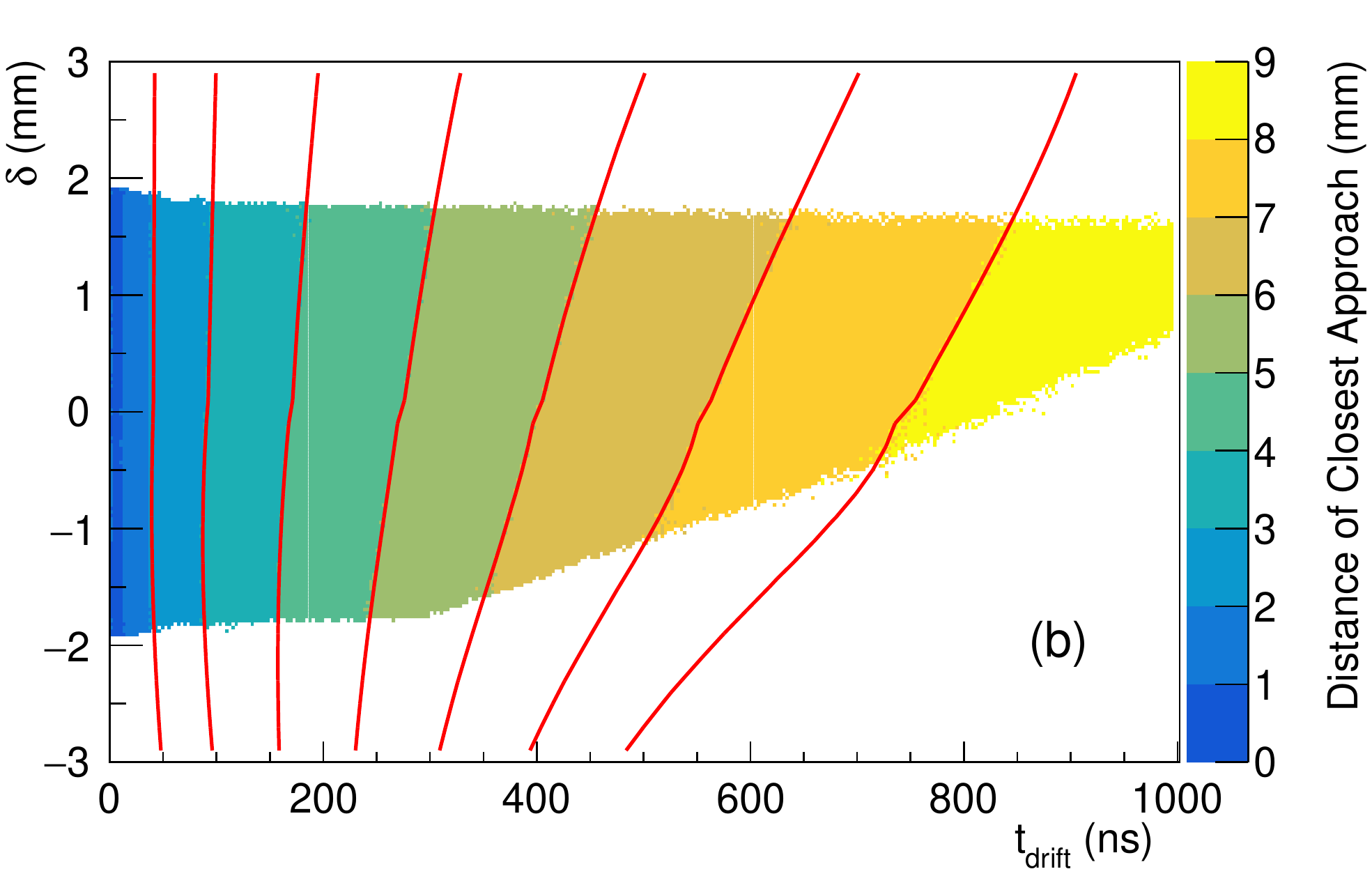}
\caption{$\delta$, the change in length of the LOCA caused by the straw deformation, is plotted against the measured drift time, $t_\mathrm{drift}$. The color scale indicates the distance of closest approach between the track and the wire, obtained from the tracking software. The red lines are contours of the time-to-distance function for constant drift distances from 2~mm to 8~mm, in steps of 1~mm. (a) shows the time-to-distance function with its initial values, obtained from simulations, and (b) shows the function after calibration, where its values match those obtained from the tracking software. 
}
\label{fig:ttod}
\end{figure}

The time-to-distance function parameters, $a_{i}$, $b_{i}$ and $c_{i}$, are obtained from a fit to data, using the tables of values for ideal straws, generated from Garfield simulations, as a starting point. An example of this process is shown in Fig.~\ref{fig:ttod}. Figure~\ref{fig:ttod}a shows the time-to-distance function before calibration. The function is drawn as a set of contour lines for drift distances in steps of 1~mm, over a plot of $\delta(z,\phi_{LOCA})$ vs drift time, with the color indicating the distance of closest approach between the track and the wire, obtained from the tracking software. The same series of drift distances was used to set the values defining the colors. 
Figure~\ref{fig:ttod}b shows the time-to-distance function after calibration. The calibration process determined the parameter values for the function and the distances obtained from the function now match those obtained from the tracking software.
After correcting for the straw distortions, the average spatial resolution improved from 180~$\mu$m to 150~$\mu$m.

\subsection{Alignment}
The calibration procedures described above assumed that the location of the CDC and the position of each wire at the two endplates were at their nominal values. An alignment calibration was performed to determine the actual location of the wires relative to the GlueX experiment.

Alignment parameters describing the difference between the designed and actual positions of the anode wires were obtained using Millepede~\cite{MillepedeManual}, a software program designed to provide an experiment-independent set of tools for detector alignment and calibrations.  Given a set of track parameters to describe the trajectory of each track, Millepede obtained the alignment parameters by iteratively minimizing the difference between the measured and predicted track parameters. The track parameters were produced by the GlueX tracking software and included 
\begin{enumerate}
	\item The residual for each measurement on the track.
	\item The uncertainty of this residual.
	\item The derivative of the residual with respect to the track parameters.
	\item The derivative of the residual with respect to the alignment parameters.
\end{enumerate}
The alignment parameters obtained for the CDC were a translation in $x$ and $y$ for each end of each wire and an offset in time (from the electronics readout) for each wire, i.e. 5 parameters for each of the 3522 wires.  The alignment was performed first for the CDC alone, and then again in combination with the Forward Drift Chambers (FDC)~\cite{Pentchev:2017omk}, allowing only rotation of the CDC with respect to the FDC for the second step. Three datasets of approximately 5 million triggers each were used: cosmic rays, beam data without magnetic field in the solenoid and beam data with magnetic field. For the cosmic data, the data acquisition was triggered by energy deposition in the barrel calorimeter alone; for the beam data, the regular production trigger was used, which requires sufficient  energy deposition in the calorimeters. 

\begin{figure}[ht]\centering
\includegraphics[width=0.45\textwidth]{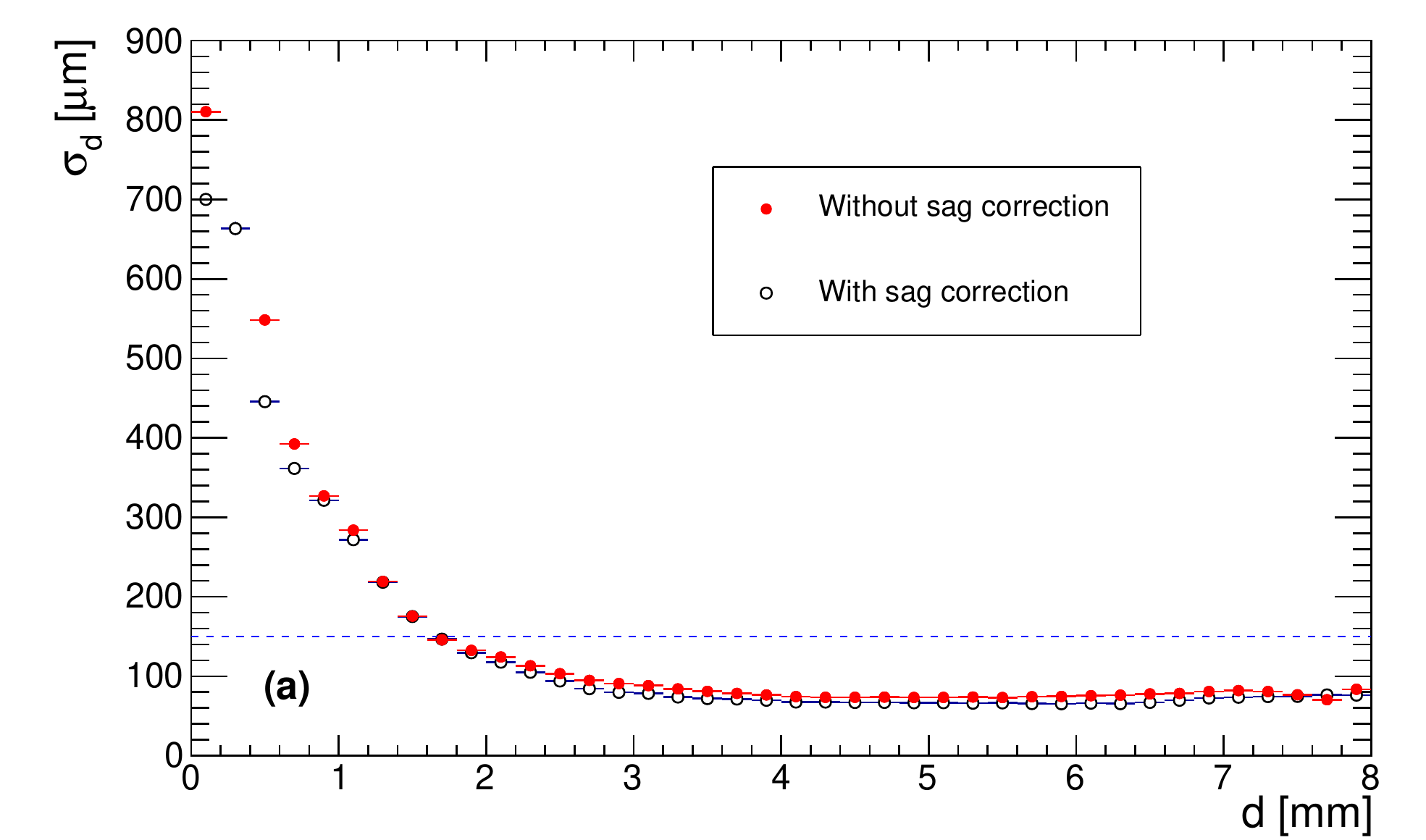}\\
\includegraphics[width=0.45\textwidth]{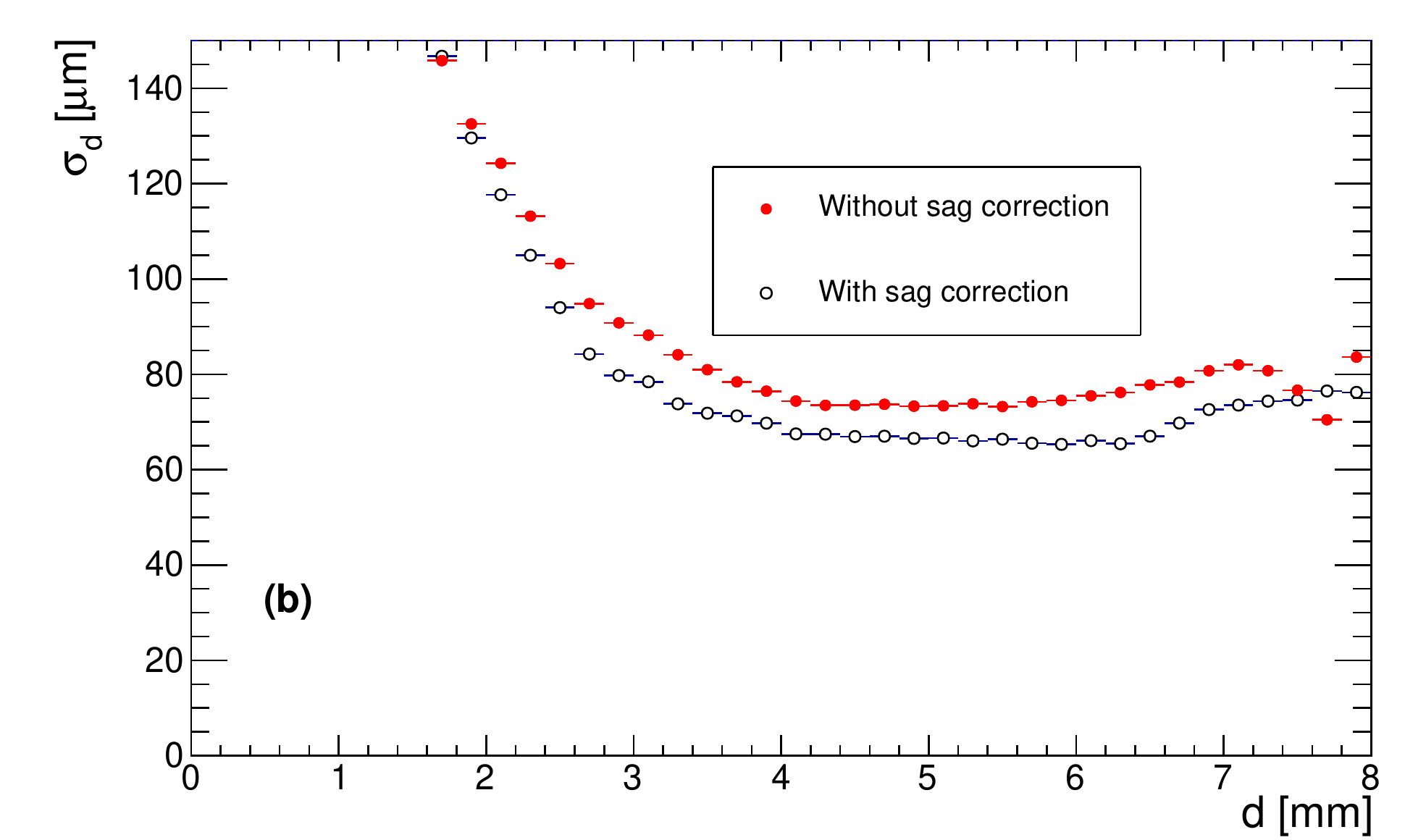}
\caption[]{\label{fig:CDCPositionResolution} Position resolution as a function of the distance of closest approach to the wire: red points are before applying the corrections for straw deflection to the drift time, black points are after applying the corrections. Figure (b) has the y-axis expanded to show the scale of the difference in the resolution.  The design resolution is shown by the dashed line at 150~$\mu$m.}
\end{figure}

The average position resolution was reduced by 10~$\mu$m as a result of the alignment process. 
The position resolution depends on the distance of closest approach of a track to the hit wire, as shown in Figure~\ref{fig:CDCPositionResolution}. Tracks passing at least 1.7~mm from the wire are measured with position resolution better than the design resolution, 150~$\mu$m. For tracks 3.5~mm or more from the wire, the position resolution is approximately 70~$\mu$m.

\section{\label{sec:tracking}Tracking and Particle Identification}
The tracking chambers were designed to reconstruct the momenta of the charged particles emerging from the target. The transverse momentum, $p_{\perp}$, and the dip--angle, $\lambda$, ($\lambda={\frac{\pi}{2}-\theta}$) are measured from the curvature of the tracks in the solenoidal magnetic field and their initial direction, with polar angle $\theta$.  The total momentum, $p$, and the longitudinal momentum, $p_{\parallel}$, are then obtained from these as $p_{\mathrm total} = p_{\perp}\sec\lambda$ and $p_{\parallel}= p_{\perp} \tan\lambda$. The accuracy of the $p_{\perp}$ measurement depends on the $r\phi$ resolution of the tracking chambers, while the $\lambda$ measurement relies on an accurate measurement of both $z$ and the distance traveled. The tracking system in the GlueX detector covers as close to a 4$\pi$ solid angle as possible over a wide range of particle momenta. The CDC has optimum coverage for polar angles between 29$^{\circ}$ and 132$^{\circ}$.  

\begin{figure}[ht]\centering
\includegraphics[width=0.45\textwidth]{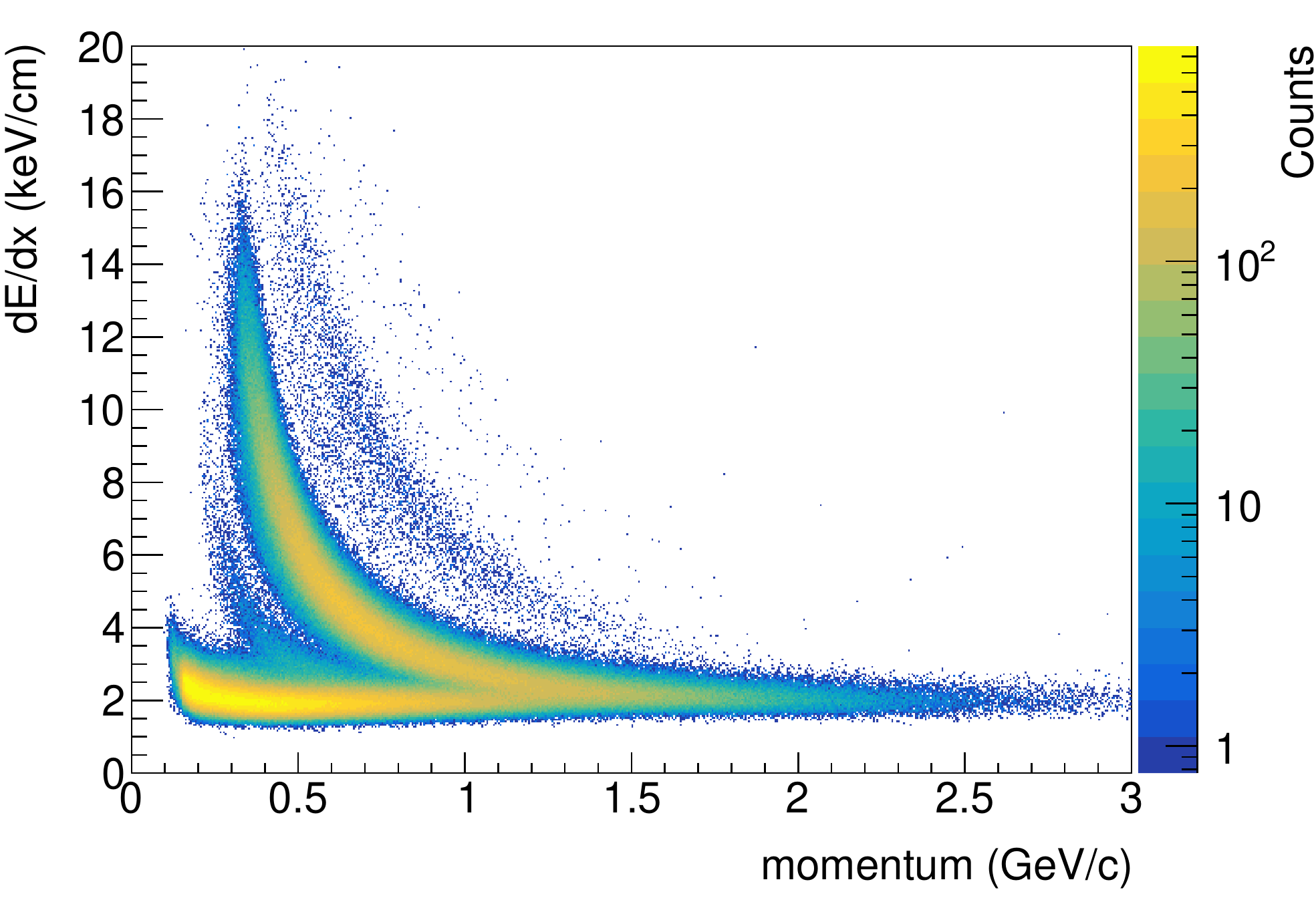}
\caption[]{\label{fig:cdc_dedx}$dE/dx$ vs momentum for positively charged particles. The proton and $\pi^+$ bands are clearly visible and the weaker K$^+$ band can be seen between them.} 
\end{figure}

The CDC provides $dE/dx$ information to aid in the separation of $\pi$'s, $K$'s and $p$'s up to momenta of about 0.45~GeV/c -- a regime where $dE/dx$ measurements work extremely well.  Fig.~\ref{fig:cdc_dedx} shows $dE/dx$ plotted as a function of momentum for positively charged particles with at least 20 straws hit per track. Proton and $\pi^+$ bands are clearly visible, and a weaker $K^+$ band can also be seen between them. $dE/dx$ for tracks in the CDC is calculated from the energy loss and track length in the straws traversed, where the energy loss is obtained from the height of the first peak of each pulse instead of its integral, as this was found to give better resolution. Pedestal subtraction is more difficult for the integral, as it suffers more from noise on the signal. This problem is greatest for tracks passing close to the wire, as the time during which the charge drifts in from all parts of the track is long (approaching 1~$\mu$s). Other contributing factors are imperfect baseline restoration after particularly large pulses, and space-charge effects, which are noticeable for tracks passing near-perpendicular to the wire: charge depletion affects the later parts of the signal, but not its initial pulse height. 

\section{Performance}
 The $dE/dx$ resolution is approximately 27\,\%, with the pion and proton bands separated by 4.3~keV/cm and 0.9~keV/cm at 0.5~GeV/c and 1.0~GeV/c, respectively. For tracks with momentum 0.5~GeV/c, the contamination under the pion and proton peaks is approximately 0\,\% and 2\,\%, respectively. For tracks with momentum 1.0~GeV/c, the contamination under the pion and proton peaks is approximately 11\,\% and 7\,\%, respectively, and at 1.5~GeV/c the two bands have merged completely. 

The detection efficiency of the CDC was studied using reconstructed particle trajectories with the solenoid on. Forward-going tracks travelled through a magnetic field from 1.6~T to 1.95~T; backward-going tracks passed through a field from 1.6~T to 1.3~T. 
The selected track sample had to fulfil certain criteria in order to limit biasing of the performance estimation. Tracks which have hits in at least 2 of both axial and stereo layers were selected to guarantee a proper definition of the direction of the trajectory. In addition, a match with the barrel calorimeter or the time of flight wall in the forward region was required. Tracks that originated outside of the target region and tracks with a large specific energy loss were not considered.

\begin{figure}[ht]\centering
  \includegraphics[width=0.45\textwidth]{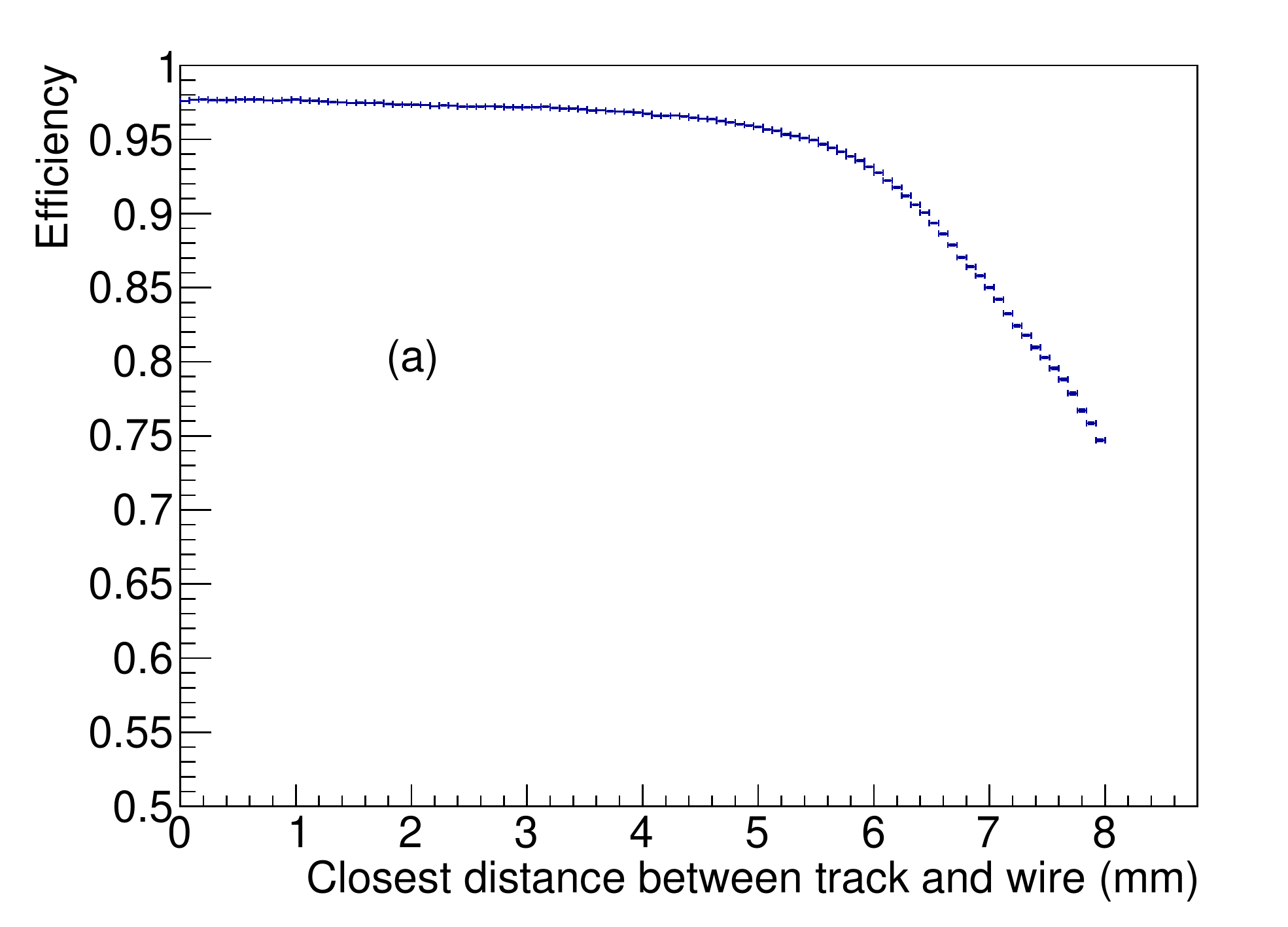}\\
  \includegraphics[width=0.45\textwidth]{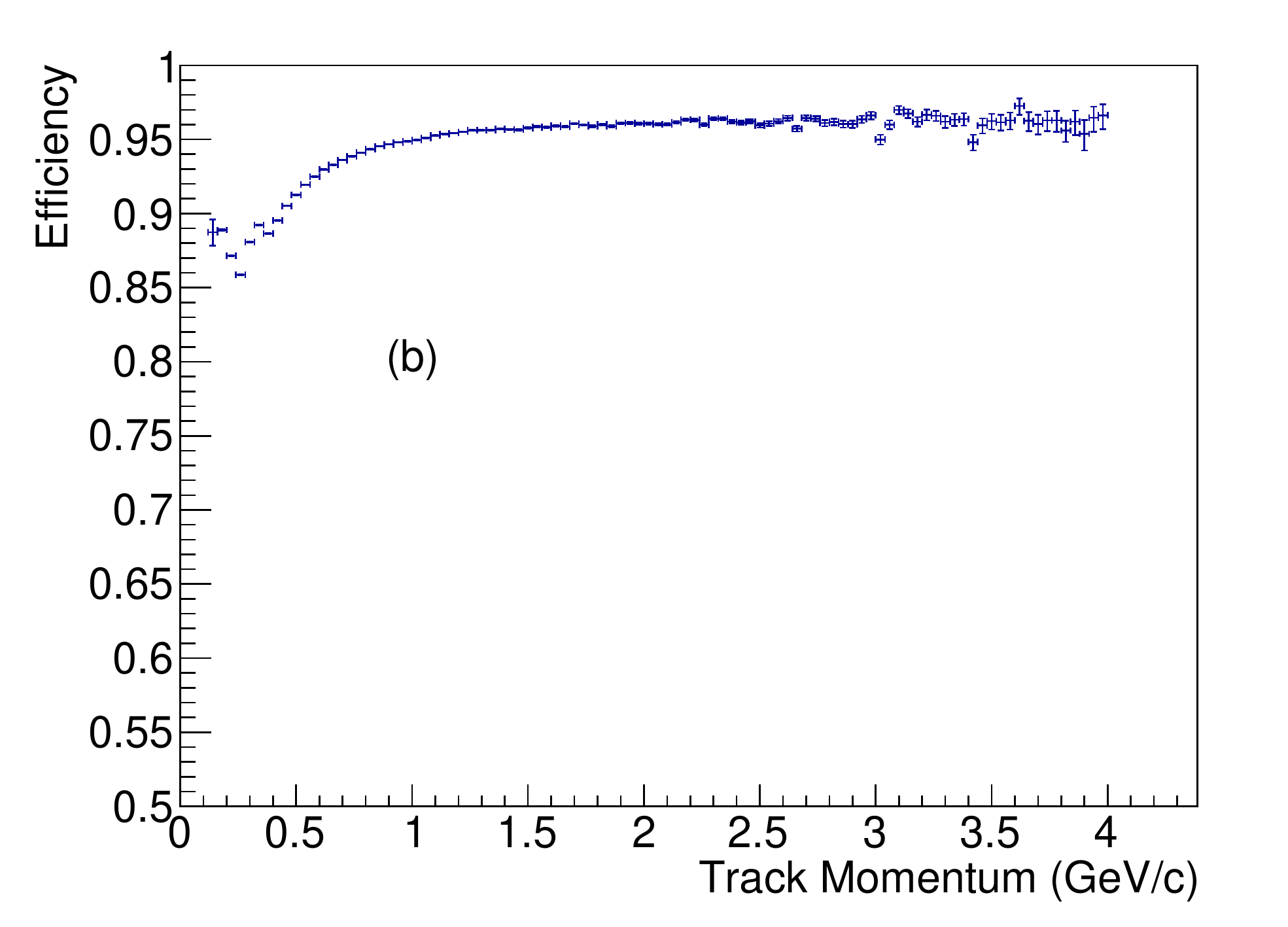}
\caption[]{\label{fig:efficiency}Detector efficiency as a function of (a) the closest distance of the track to the wire and (b) the track momentum.} 
\end{figure}

Figure~\ref{fig:efficiency}a shows the mean detection efficiency for all channels in the CDC, measured with the solenoid on. 
The average efficiency is above 98\,\% for tracks that traverse the straw-tubes within 4~mm of the sense wire. The efficiency decreases towards the walls of the tube, as the track length and signals become smaller and the drift time becomes longer, which increases the chance of a noise signal arriving first and obscuring the real hit. This does not pose a problem for the track reconstruction, since the close-packing of the tubes ensures that tracks passing through the low efficiency region of one tube also pass through the high efficiency region of the tubes in the surrounding layers. The detector shows a very uniform efficiency above 96\,\% for tracks with momenta between 1 and 4~GeV/c (see Fig.~\ref{fig:efficiency}b). For lower momenta, short particle trajectories and multiple scattering complicate the clean track definition necessary for a proper evaluation of the performance.

\begin{figure}[ht]\centering
\includegraphics[width=0.45\textwidth]{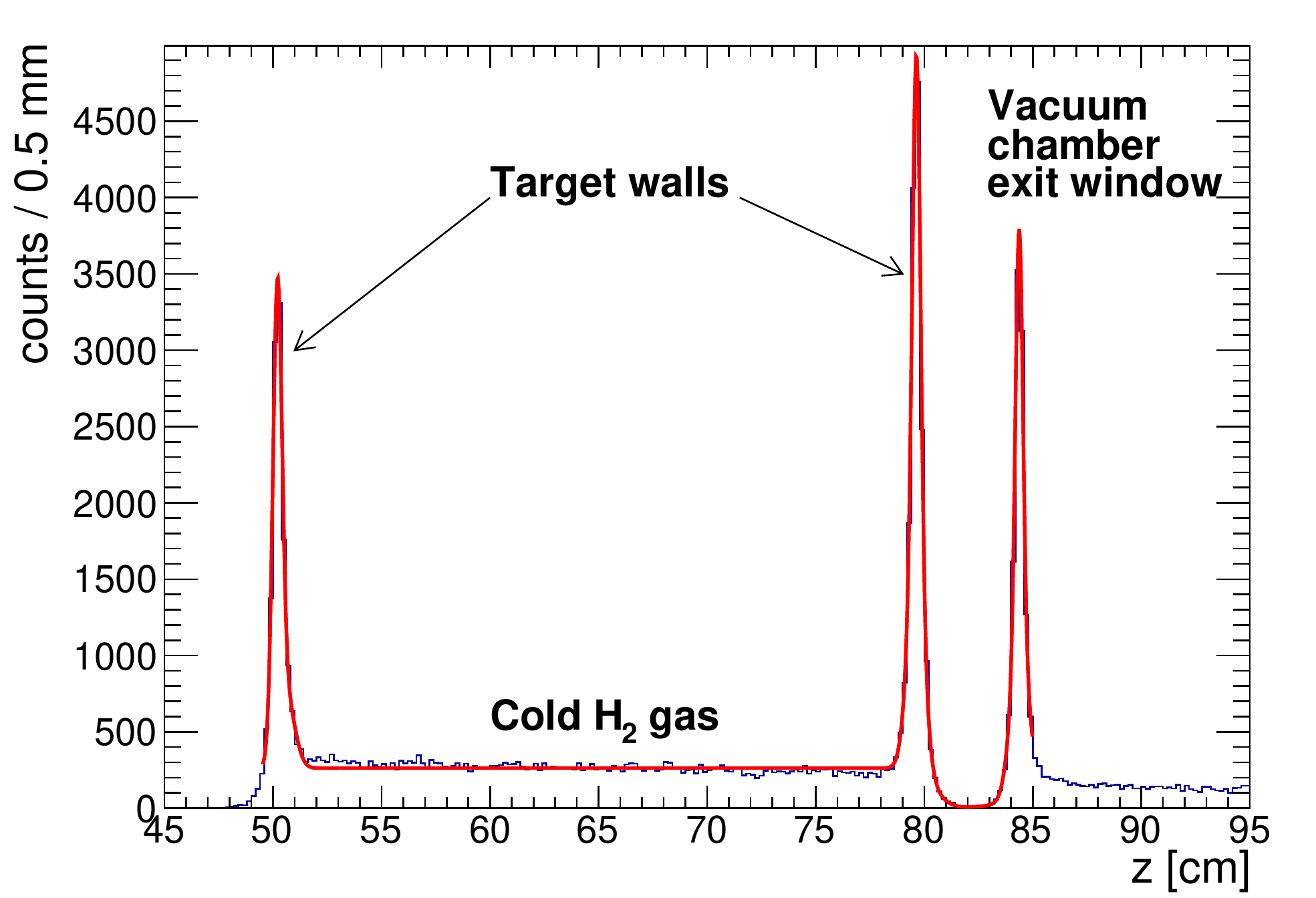}
\caption[]{\label{fig:zvertex}Vertex position near the beam line for an empty target run with the solenoid on, reconstructed from pairs of tracks.  The vertex was required to be within 5~mm radial distance with respect to the beam line.}
\end{figure}

The walls of the cryogenic target and the exit window for the vacuum chamber surrounding the target are sufficiently thin that they can be used to estimate the position resolution of the
detector.  Figure~\ref{fig:zvertex} shows the vertex position along the beam line using pairs 
of tracks reconstructed in the CDC for an empty target run.  The three peaks in the figure yield resolutions between 2.4~mm and 2.8~mm. From this, the estimated $z$-vertex resolution is 2.6~mm. 

\begin{figure}[ht]\centering
\includegraphics[width=0.45\textwidth]{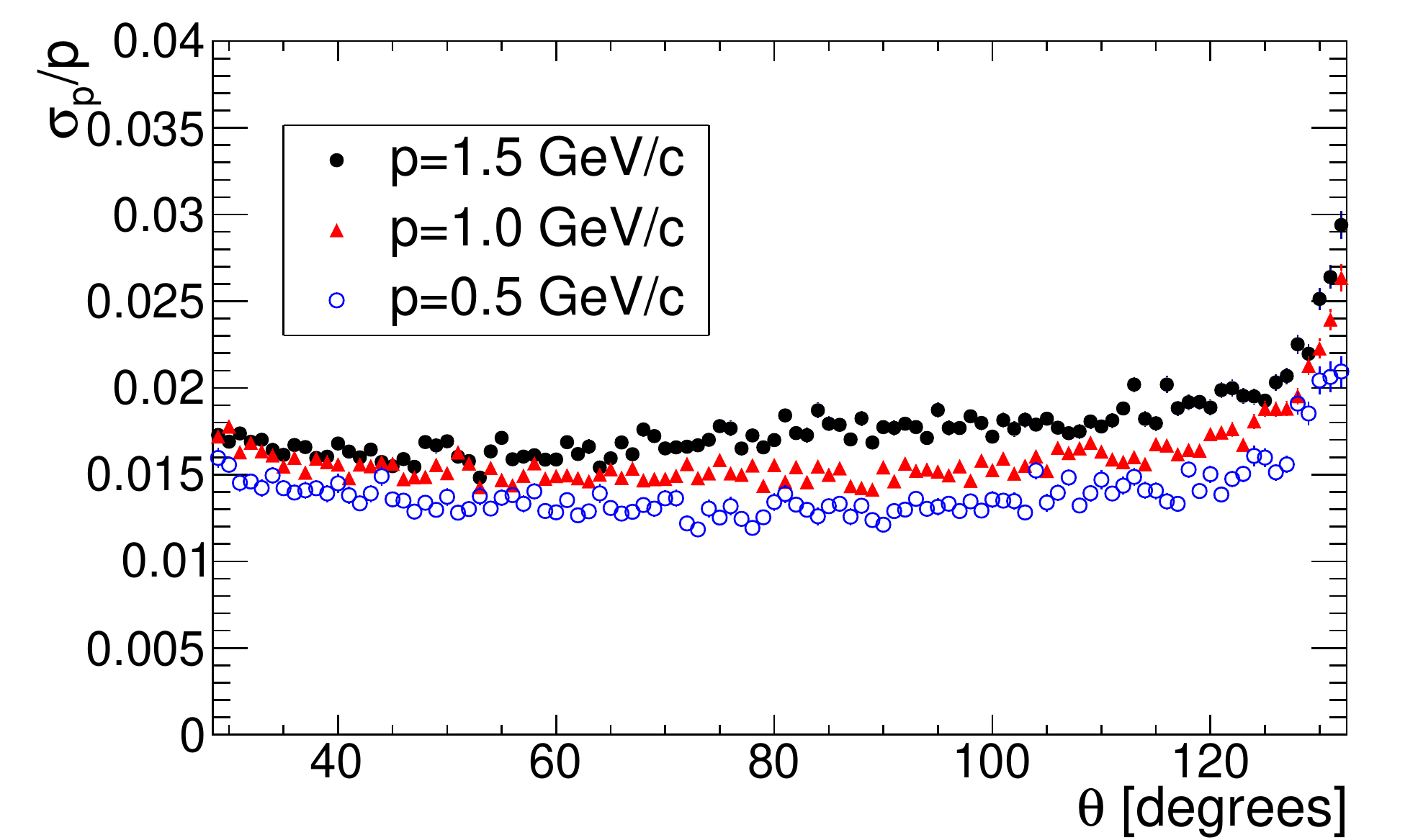}
\caption[]{\label{fig:momentum_resolution}Momentum resolution $\sigma_p/p$ for simulated pion tracks, as a function of polar angle, $\theta$, and momentum.  The range in $\theta$ is 29--132$^\circ$.}
\end{figure}

\begin{figure}[ht]\centering
\includegraphics[width=0.45\textwidth]{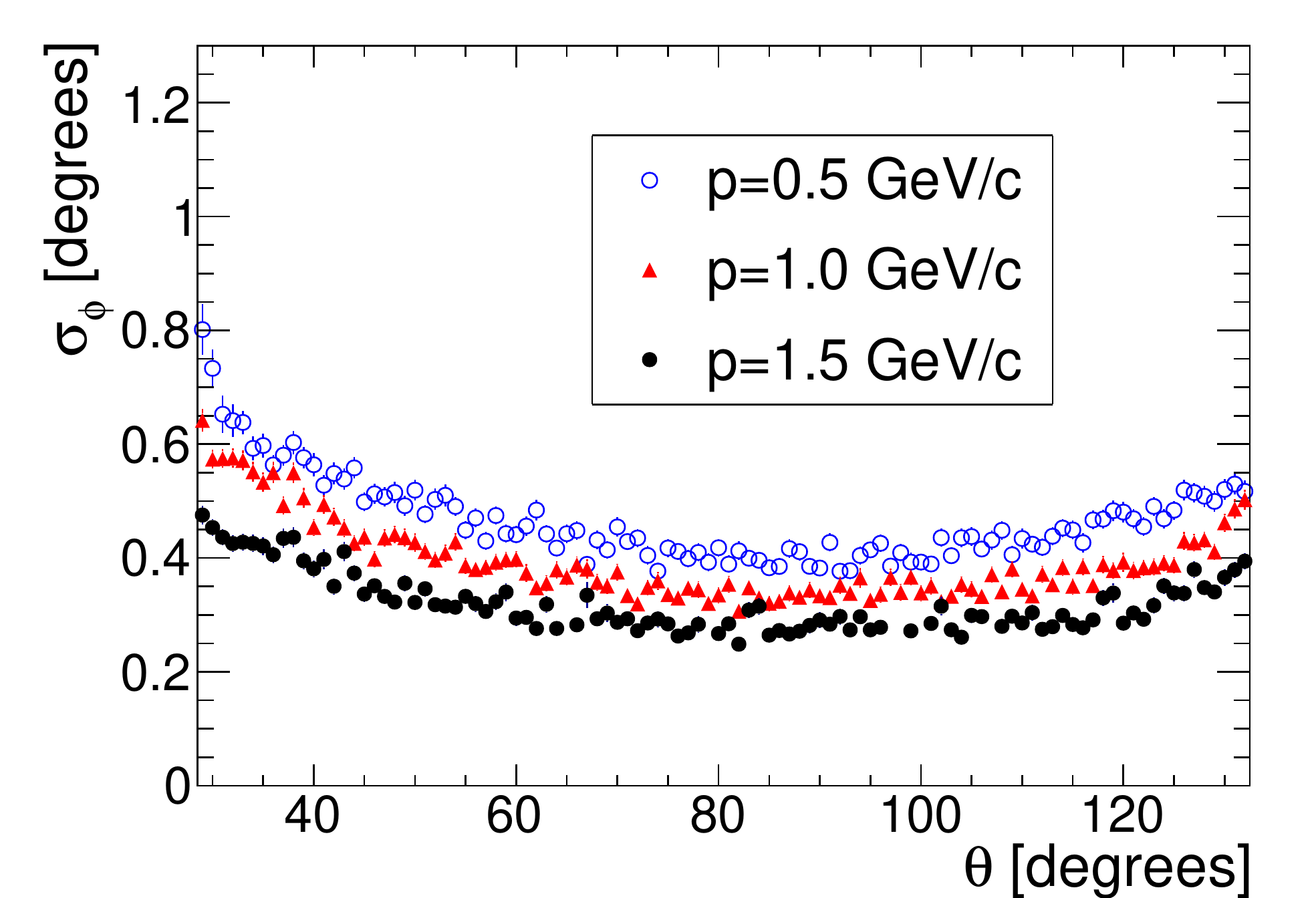}
\caption[]{\label{fig:phi resolution}Azimuthal angle resolution, $\sigma_\phi$, for simulated pion tracks, as a function of polar angle, $\theta$, and momentum.  The range in $\theta$ is 29--132$^\circ$.}
\end{figure}

\begin{figure}[ht]\centering
\includegraphics[width=0.45\textwidth]{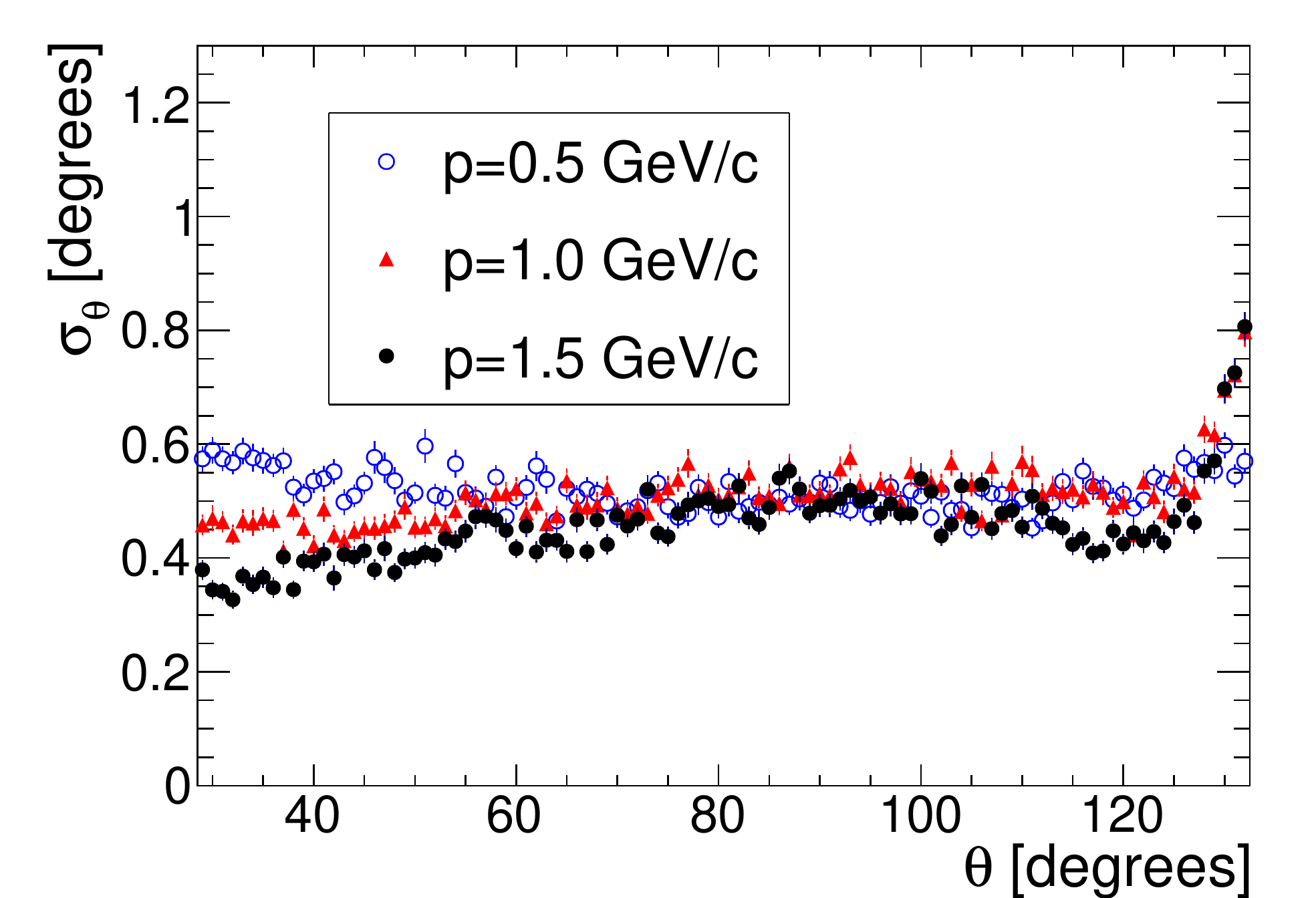}
\caption[]{\label{fig:theta resolution}Polar angle resolution, $\sigma_\theta$, for simulated pion tracks, as a function of polar angle, $\theta$, and momentum. The range in $\theta$ is 29--132$^\circ$. }
\end{figure}

Due to the technical difficulties of determining the momentum resolution from data, a simulation was used to estimate the momentum resolution of the CDC.  
The simulation incorporated a realistic model of the material and geometry of the detector and the timing resolution was tuned to agree with the measured values. Single pion tracks with discrete values of the magnitude of the momentum were thrown from the target region in the range 500--800~mm in $z$, with the center of the target at z=650~mm.
The momentum resolution is shown in Fig.~\ref{fig:momentum_resolution} as a function of polar angle for 0.5, 1.0, and 1.5~GeV/c pions.  The momentum resolution is better than the 2\,\% goal for most of the angular coverage of the CDC.  The angular resolution for these three momenta is shown in Fig.~\ref{fig:phi resolution}, for azimuthal angle, and Fig.~\ref{fig:theta resolution}, for polar angle.

\section{Summary}
The GlueX CDC was built between 2010 and 2013. It was installed in the GlueX experiment in 2014, and has been operated since that time. The drift chamber has exceeded its $r\phi$ resolution goal of 150~$\mu$m, and has high reconstruction efficiency for individual hits in the detector. 

Sagging of many of the straws was discovered after installation, but the cause of this is not fully understood. A calibration procedure was developed to handle these distortions and it was found that they do not impact the chamber performance significantly. 

Under normal GlueX running conditions, the currents on all straws in the chamber remain at safe levels.  During its five years of operation, only 2 of the straw-tubes have stopped working, due to their wires breaking within the first year. To date, the remaining 3520 straw-tubes are still in use, with no signs of aging. 

\section{Acknowledgements}
We wish to thank Mason Blaschak, Madison Brumbaugh, Tom Charley, Brent Driscoll, Ariana Golden, Liz Keller, Rahul Kurl, Devin McGuire, Kaitlin Mueller, Aleksandar Popstefanija, Gary Wilkin and Amy Woodhall for their tireless work in building the CDC. The Carnegie Mellon Group is supported by the U.S. Department of Energy, Office of Science, Office of Nuclear Physics, DOE Grant No. DE-FG02-87ER40315. This material is also based upon work supported by the U.S. Department of Energy, Office of Science, Office of Nuclear Physics under contract DE-AC05-06OR23177.

%
%

%
\end{document}